%% file: paper.tex
\documentclass[]{aa}
\usepackage[utf8]{inputenc}
\usepackage[bookmarks=false]{hyperref}
\usepackage{subcaption}
\usepackage{natbib}
\usepackage{aas_macros}
\usepackage{aalongtable}
\usepackage{mathtools}
\usepackage{bm}

\begin{document}

\title{Decoherence in LOFAR-VLBI Beamforming}
\titlerunning{Decoherence in LOFAR-VLBI Beamforming}

\author{Etienne Bonnassieux$^{1,2}$, Alastair Edge$^{3}$, Leah Morabito$^{3}$, Annalisa Bonafede$^{1,2}$}

\institute{
Universita di Bologna, Via Zamboni, 33, 40126 Bologna BO, Italy
\and
INAF, Instituto di Radioastronomia, Via Piero Gobetti, 101, 40129 Bologna BO, Italy
\and
Centre for Extragalactic Astronomy, Department of Physics, Durham University, Durham DH1 3LE, UK
}

\authorrunning{E. Bonnassieux, A. Edge, L. Morabito, A. Bonafede}

\input{source/Defs.tex}

\input{source/abstract.tex}

   \maketitle


\input{source/intro.tex}

\input{source/sec0.tex}

\input{source/sec1.tex}

\input{source/sec2.tex}
\input{source/sec3.tex}

\input{source/sec4.tex}

\input{source/conclusion.tex}

\begin{acknowledgements}
The authors thank the LOFAR Surveys KSP Primary Investigators for allowing us to use their data for our tests and images in this paper. We also wish to thank Philippe Zarka for his help in seeing the relevance of this paper in the context of NenuFAR's "international station" mode. Etienne Bonnassieux \& Annalisa Bonafede acknowledge support from the ERC-ERG grant DRANOEL, n.714245. ACE acknowledges support from STFC grant ST/P00541/1. 
\end{acknowledgements}

\bibliographystyle{aa}
\bibliography{source/bib}

\end{document}

%% file: source/Defs.tex
\def\pg{\paragraph{}}

\def\weight#1{\omega_{#1}}
\def\weightpr#1{\omega'_{#1}}
\def\vecweight#1{\bm{\omega}_{#1}}
\def\matJ#1{\bm{{J}}_{#1}^{t\nu}}
\def\lvec{\bm{l}}
\def\uvec#1{\bm{u}_{#1}}
\def\matK#1{\bm{{K}}_{#1,\lvec}^{t\nu}}
\def\matAa#1{{{A}}_{#1}^{t\nu}}
\def\matA#1{{{A}}_{#1,\lvec}^{t\nu}}
\def\matAo#1{{{A}}_{#1,0}^{t\nu}}
\def\matAlo#1{{{A}}_{#1,\lvec_0}^{t\nu}}
\def\k#1{k_{#1,\lvec}^{t\nu}}
\def\ko#1{k_{#1,0}^{t\nu}}
\def\klo#1{k_{#1,\lvec_0}^{t\nu}}
\def\matV#1{\bm{V}_{#1}^{t\nu}}
\def\matVhat#1{\hat{\bm{V}}_{#1}^{t\nu}}
\def\matB{\bm{B}}
\def\matBl{{\matB}_{\lvec}^{\nu}}
\def\matBlo{{\matB}_{\lvec_0}^{\nu}}
\def\I{\mathbf{I}}
\def\dirtyIpq{I_{\text{dirty}}^{pq,t\nu}}
\def\dirtyIpqsmr{\tilde{I}_{\text{dirty}}^{pq,t\nu}}
\def\dirtyI{I_{\text{dirty}}}

%% file: source/abstract.tex

\abstract{
We show that the use of a superstation (a phased array created using multiple stations of an interferometric array) created in post-processing for LOFAR-VLBI observations introduces a direction-dependent loss of signal in the image. We show this effect using simulations and real data. Using the RIME formalism, we characterise it fully, and give limits under which this signal loss is negligible. Finally, we show that we are able to fully predict this effect. We close with guidelines for interferometric observers to avoid this effect in their observations, and a discussion of techniques which could limit this effect or do away with it entirely. The latter in particular will be relevant to the SKA should its long baselines be used to their fullest potential.
}

%% file: source/intro.tex

\section{Introduction}\label{sec.intro}

\pg
The new generation of Square Kilometer Array (SKA) pathfinder instruments 
is pushing the boundaries of what was previously achievable with radio interferometers. The LOw Frequency ARray LOFAR \citep{2013A&A...556A...2V}, in particular, is in full science production mode, mapping the sky at very low frequencies and high resolution. However, the instrument is not yet used to its fullest extent as a matter of course: in particular, the use of LOFAR-VLBI (i.e. LOFAR with its ``international", or European, stations - as opposed to using only those stations in the Netherlands) is still an ongoing and active area of technical development, with great strides still being made \citep[e.g.][and a paper splash scheduled for next year]{2016A&A...595A..86J,2015A&A...574A.114V,2016yCat..35930086V,2016MNRAS.461.2676M,2019A&A...631A..49K}.

\pg
One {aspect} of interferometric technique in the LOFAR-VLBI regime is that it blurs the conceptual boundaries between ``traditional" interferometry and Very Long Baseline Interferometry (henceforth VLBI). Historically, {although} the two disciplines share common ancestry and common techniques, the specific case of VLBI allowed for a set of simplifications to the general problem of interferometry, at the cost of certain constraints and limitations. In particular, VLBI focused on small fields of view (FoV), extremely high resolutions, and single-object observations. This freed it from the direction-dependent effects which require correcting for modern ``standard" interferometric observations, such as Dutch LOFAR's Two-meter Sky Survey (LOTSS) fields \citep{2019A&A...622A...1S}. Since VLBI focuses on single-object pointings, it also did not need to account for effects such as time/frequency smearing \citep[see][and companion papers]{2011A&A...527A.106S}, which increase signal loss as a function of angular distance from phasing centre - typically very small angular distances for VLBI observations.

\pg
LOFAR-VLBI aims to use the full LOFAR array, which includes baselines over 1000km long \citep{2013A&A...556A...2V} and therefore lies squarely within the limit of VLBI as understood from an instrumental perspective. However, it also aims to take advantage of the tremendous opportunities offered by the LOFAR-VLBI Field of View (FoV) of $1.15^\circ$ \citep[{FWHM of the primary beam quoted in}][]{2013A&A...556A...2V} - allowing for multiple objects to be detected at once, at sub-arcsecond angular resolutions. This places LOFAR-VLBI outside of the scope of ``traditional" VLBI, as it introduces additional constraints: we now have to properly account for direction-dependent effects and smearing if we want to take advantage of the wide field of view by imaging away from phase centre. 

\pg
In many cases, one bottleneck to LOFAR data reduction is its sheer size. As such, the LOFAR-VLBI pipeline performs post-processing beamforming of all the core LOFAR stations (i.e. {the 24 LOFAR stations closest to the central core}). This results in an 80\% reduction in data size, making its reduction much more tractable. It also increases the sensitivity of the baselines pointing from this superstation to the international stations by a factor of $\sqrt{N}$ with N the number of stations being beamformed, which is necessary to calibrate the long baselines. However, this comes at a cost: should the astronomer's science target be somewhat extended, such a drastic superstation formation will inevitably result in direction-dependent flux suppression. For example, the exact amount of suppression resulting from using all core LOFAR stations as a single superstation (which is standard practice with LOFAR-VLBI at the time of writing) is given later in the paper, in Fig. \ref{fig.ldecorr}. This figure is referred out of order so as to preserve the paper's throughline, and thus its clarity.

\pg
For most science cases, the source of interest may either not be at phase centre or may be mildly extended. In such cases, even $30''$ away from phase centre - {which, to give an idea of scale}, corresponds to 6 pixels in LOTSS images\footnote{{LOTSS is not affected by this decoherence effect, and is used only as a benchmark for pixel size.}} \citep{2019A&A...622A...1S} - scientists can expect an extinction factor of about 40\%. This will have major effects on the fidelity of their source modeling, on the quality of their calibration solutions, and on the reliability and the veracity of their scientific work. Worse, this may not be immediately apparent to the astronomer. 

\pg
The aim of this paper is to explain the source of this flux suppression, the regime in which it occurs, and how to avoid it. Though much of the discussion here is quite technical and targets a specialist audience, we have taken care to make it as accessible as possible to astronomers by explaining the practical consequences of our results throughout the paper. 



\pg
The paper structure is as follows: in Sections 2 \& 3, the scientific context for post-processing superstation formation will be discussed. In Section 4, a full mathematical development of this technique and its effect on interferometric measurement is given. This development is the cornerstone of this paper, and the foundation from which we build our analysis. In Section 5, we show that we not only correctly model the way current suites of LOFAR data reductions perform post-processing superstation formation, but that our analytical conclusions are valid; this is done by creating a simulated dataset. In Section 6, we perform the same analysis using real data. We close with a general discussion and practical guidelines for astronomers on how to avoid the issue outlined in this paper.

%% file: source/sec0.tex

\section{LOFAR, VLBI, and Superstation Beamforming}\label{sec.0}

\pg
Good telescopes are designed for their resolution to be diffraction-limited. The very long wavelength of radio frequencies means that, to obtain resolutions comparable to what is routinely achieved in the optical regime, extremely large antennas are necessary. Historically, this has led to two relatively divergent paths: monolithic antennas (e.g. the Effelsberg 100m Radio Telescope, Green Bank Telescope) and aperture synthesis (which can be further broken down into phased arrays and interferometric arrays). For the remainder of this paper, we consider only the second family of instruments.

\pg
The difference between the two types of aperture synthesis array lies in the way in which signal from their constituent antennas is combined: phased arrays rely on a sum operation (and are hence also known as $\sum$-arrays) while interferometric arrays rely on a multiplication operation (and are hence also known as $\Pi$-arrays). 
LOFAR is exceptional (but not unique) in that it is an interferometer made up of phased arrays. This allows it to reach very high sensitivity and resolution at a low cost.



\pg
Very Long Baseline Interferometry (henceforth VLBI) is an interferometric technique which consists of combining the signal from multiple telescopes at very large distances from each other, forming baselines greater than $10^6$ wavelengths in length as a single interferometer. This introduces technical constraints on observations (e.g. very accurate clocks needed to combine the signals from different instruments properly (and offline), extremely sparse $uv$-coverage, geometric correlator model needs high accuracy) but is an extremely powerful method to reach very high resolutions in instances where the diffraction limit of the component stations would otherwise forbid it, by creating a sparse interferometer with them. 

\pg
At the very high spatial frequencies (i.e. very high resolutions) that the instrument probes, the morphological structure of sources often become resolved and there is no guarantee that much signal is present in a given baseline. Consequently, VLBI calibration is often limited by the signal-to-noise ratio (henceforth SNR) of a given observation. Since signal calibration effects are functions of time and frequency, one approach sometimes used to improve SNR and therefore calibration is the post-processing beamforming of nearby central antennas, with the benefits outlined above (improvement of $\sqrt{N}$ in sensitivity, with N the number of antennas combined into a superstation). Using central stations, rather than outlying ones, helps further constrain calibration by increasing sensitivity on similar spatial frequencies multiple times.

\pg
If the distance between these central antennas is sufficiently small compared to the distance between these antennas and others, then instead of probing N slightly different spatial frequencies once (with N the number of central antennas), these individual samples can be approximated as probing the same spatial frequency N times, thereby decreasing the noise associated with that spatial frequency by a factor of $\sqrt{N}$. The question then becomes: for LOFAR, under what limits can this be done, and what constraints does this technique impose?

%% file: source/sec1.tex

\section{Decoherent Phasing in LOFAR Beamforming}\label{sec.1}

\pg
The main difference between standard VLBI superstation forming and LOFAR is that, in VLBI, one has access to the voltage information (i.e. analog signal) of individual stations prior to visibility formation, whereas in the case of LOFAR the superstation formation is done after visibility formation (i.e. we only have access to the correlated signal). This is why we refer specifically to \textit{post-processing} superstation beamforming in the case of LOFAR. In this section, we will highlight the problem this introduces in terms of signal coherence - which is the quantity actually measured by interferometers.

\pg
Let us begin by considering the case of traditional (pre-processing) superstation beamforming from the antenna voltages. This simply reduces to creating a phased array using the stations being beamformed: a phase factor is applied to the voltages so as to have fully coherent combination at the phase centre, the voltages are then averaged together, and this new voltage is then used to form visibilities with other stations. In other words, the order of operations is as follows:
\begin{align}
v_r    &= \frac{1}{N_R}\sum_{p \in R}^{N_R} v_p e^{2\pi i \phi_{p}}\\
V_{rq} &= \frac{1}{N_R}\langle v_r v_q^* \rangle_{\delta t,\delta \nu} \label{eq.vis.form}\\
       &= \frac{1}{N_R}\langle \sum_{p \in R}^{N_R} v_p v_q^* e^{2\pi i \phi_p}  \rangle_{\delta t,\delta \nu} \label{eq.vis.1}
\end{align}
where $r$ is the index of the superstation being formed, $v_p$ is the voltage measurement associated with antenna $p$, $R$ is the set of $N_R$ antennas used to form $r$, $q\notin R$ is some remote antenna, $\phi_p$ is the phase correction for antenna $p$, and $\delta t, \delta\nu$ are the correlator dump time and channel bandwidth respectively. We assume here that all visibilities have equal weight. Eq. \ref{eq.vis.form} is the equation for the formation of visibilities from antenna voltages: {it is} this operation that the correlator performs. The point here is that, in this case, the sum over $p$ is done prior to correlation. 

%

\pg
In post-processing, however, we no longer have access to the information prior to correlation. {For LOFAR, this means that we no longer have access to the voltage information of the antennas which constitute each station, but only to the station-level information: the beamforming is done with each station acting as a single antenna, rather than by phasing up every physical antenna in the set of stations being beamformed.} As such, the phasing must be done from visibilities:
\begin{align}
V'_{rq} &= \frac{1}{N_R}\sum_{p \in R} V_{pq}\\
       &= \frac{1}{N_R}\sum_{p \in R} \langle v_p v_q^* e^{2\pi i \phi_p} \rangle_{\delta t,\delta \nu} \label{eq.vis.2}
\end{align}
where $V'_{rq}$ is the post-processing estimate of $V_{rq}$. As we can see, the sum over $p$ in Eq. \ref{eq.vis.1} is performed before averaging over time and frequency, while it is performed after the fact in Eq. \ref{eq.vis.2}. In other words, this summation is done \textit{before correlation} in the first case, and \textit{afterwards} in the second. These two operations will only be equivalent if the variables being correlated are independent random variables. If not, some amount signal coherence will inevitably be lost.


\pg
It should be noted that this effect - or a similar one - has been noticed over the course of developing the LOFAR-VLBI pipeline by its working group (Morabito, pers. comm.). However, this effect was not fully understood nor modeled at the time. As such, understanding it - and thus mitigating it - is of immediate interest: by characterising it fully, we can ensure that it does not bias our existing pipelines \& associated results.

\pg
The mathematical framework used thus far, however, is not well-suited to analysing this problem in depth. For this reason, we will pursue the same line of reasoning using a more appropriate framework: the Radio Interferometer's {Measurement} Equation, or RIME.

%

%% file: source/sec2.tex

\section{Analytical Framework: the RIME approach}\label{sec.2}

\pg
In this section, we give an analytical development of what post-processing superstation beamforming entails using the Radio Interferometer's Measurement Equation \citep[see][and references therein]{2011A&A...527A.106S,2011A&A...527A.107S,2011A&A...527A.108S,2011A&A...531A.159S}. This allows us to find a quantitative estimate for a baseline-dependent error factor, and thus to predict the overall decoherence introduced by this operation for a point source at a given distance from phase centre.

\subsection{Predicting Decoherence for 1 Visibility}

\pg
Let us begin by writing what the operation of superstation beamforming entails by considering the creation of 1 superstation visibility pointing to antenna $q$ from superstation $r$, formed with a set of antennas $p$. We allow for the presence of weights. The $uvw$ coordinates and weights of our superstation are thus the following:
\begin{align}
\uvec{rq} = \begin{pmatrix} u_{rq}\\ v_{rq}\\ w_{rq}\\ \end{pmatrix} &= \frac{1}{\displaystyle\sum_{p\in \text{ST}} \weight{pq}}
\begin{pmatrix} \displaystyle\sum_{p\in \text{ST}} \weight{pq} u_{pq} \\
\displaystyle\sum_{p\in \text{ST}} \weight{pq} v_{pq} \\
\displaystyle\sum_{p\in \text{ST}} \weight{pq} w_{pq}
\end{pmatrix} \label{eq.STuvw} \\
\weight{rq} &= \displaystyle\sum_{p\in \text{ST}} \weight{pq}\label{eq.weights}
\end{align}
where $\text{ST}$ is the set of stations which are being beamformed into the superstation, $(u_{pq},v_{pq},w_{pq})$ are the $uvw$-coordinates associated with visibility $V_{pq}$, and $\weight{pq}$ is the associated weight, accounting for data flagging. Note that, in the measurement sets used in this paper, the averaged $uvw$ coordinates of Eq. \ref{eq.STuvw} are actually stored as $(u_{qr},v_{qr},w_{qr})$ because the new visibilities are added {after} the existing visibilities, and not before. This introduces a factor of $-1$ to the stored visibilities, since $\bm{u}_{rq} = -\bm{u}_{qr}$. This has no impact on imaging, and thus only appears within the dataset. The weights $\weight{rq}$ are unaffected.

\pg
The calculated visibilities of our beamformed superstation, meanwhile, are given by:
\begin{align}
\matVhat{rq} &= \frac{1}{\displaystyle\sum_{p\in \text{ST}} \weight{pq}} \left(\displaystyle\sum_{p\in \text{ST}} \weight{pq} \matV{pq}  \right)\label{eq.vis.st}
\end{align}
i.e. the visibility between superstation $r$ and antenna $q$ is simply the weighted average of all visibilities from antennas used to form the superstation and antenna $q$. The hat denotes that this is an estimation. $\matV{rq}$ is a $2 \times 2$ complex matrix. Using the RIME, we can write $\matV{pq}$ as:
\begin{align}
\matV{pq} &= \sum_{\lvec} \left(\matJ{p,\lvec} \matK{p} \matBl \left(\matK{q}\right)^T \left(\matJ{q,\lvec}\right)^T \right)\label{eq.vis}
\end{align}
where $\matBl$ is the brightness distribution matrix at position $\lvec=(l,m,n-1)$ (where $(l,m,n)$ are cardinal angles and we make the small-angle approximation that $n=\sqrt{1-l^2-m^2}$) and frequency $\nu$, $\matK{p}$ is a scalar Jones matrix which encodes which point in $uv$-space antenna $p$ contributes to sampling, and $\matJ{p,\lvec}$ is the Jones matrix associated with antenna $p$ and direction $\lvec$. This Jones matrix encodes the propagation effects which affect the signal as it travels between its source, located at $\lvec$, and the final measurement at the end of the instrumental chain. All of these matrices are complex $2 \times 2$ matrices. $\matK{p}$ and $\matK{q}$ are scalar matrices:
\begin{align}
\matK{p} &= \I e^{-2\pi i \uvec{p} \cdot \lvec} = \I\k{p}
\end{align}
and they therefore commute with all other Jones matrices.

\pg
Let us now make a simplifying hypothesis, and assume that we are unaffected by propagation signals: $\matJ{p,\lvec} = \I \forall ( \lvec,p,t,\nu )$, where $\I$ is the $2\times 2$ unit matrix. This is equivalent to assuming that we have perfectly corrected all calibration effects, including direction-dependent ones. In this limit, Eq. \ref{eq.vis} can be written as:
\begin{align}
\matV{pq} &= \sum_{\lvec} \left(\matBl \k{pq}  \right)\label{eq.vis.simpl}\\
\k{pq}    &= \k{p}(\k{q})^* =  e^{2\pi i \left(\uvec{p}-\uvec{q}\right) \cdot \lvec}
\end{align}
where $\k{pq}$ now becomes the Fourier kernel which determines the mapping between the brightness matrix and the visibility. We can now rewrite Eq. \ref{eq.vis.st} as:
\begin{align}
\matVhat{rq} &= \frac{1}{\displaystyle\sum_{p\in \text{ST}} \weight{pq}} \left(\displaystyle\sum_{p\in \text{ST}} \weight{pq} \sum_{\lvec} \left(\matBl \k{pq}  \right) \right)\\
          &=  \sum_{\lvec}\frac{1}{\displaystyle\sum_{p\in \text{ST}} \weight{pq}} \left(\displaystyle\sum_{p\in \text{ST}} \weight{pq} \left(\matBl \k{pq}  \right) \right)\label{eq.vis.st.simp}
\end{align} 

\pg
Within the limit of perfect calibration, we can also analytically predict the expected value of the superstation visibility $\matV{rq}$, since we know its associated $\uvec{rq}$ coordinates exactly. Indeed, with the definitions above, we can show straightforwardly that it can be written as:
\begin{align}
\matV{rq} &= \sum_{\lvec} \left( \matBl \k{rq} \right) \label{eq.vis.st.exact}
\end{align} 

\pg
If Eq. \ref{eq.vis.st.simp} is exact, then it should give the same result as \ref{eq.vis.st.exact}. We thus equate them with a proportionality factor. If this proportionality factor is unity, then Eq. \ref{eq.vis.st.simp} is exact. Otherwise, this proportionality factor gives us an indication of some error factor introduced by the post-processing superstation beamforming. Let us define our proportionality factor as $A$. We will assume it is a scalar and function of the same parameters as $\matV{rq}$, and therefore a function of $(r,q,t,\nu)$ at least. We can now write:

\begin{align}
\matAa{rq} \matV{rq} &= \matVhat{rq}\\
\matAa{rq} \sum_{\lvec} \left( \matBl \k{rq} \right) &= \sum_{\lvec}\frac{1}{\displaystyle\sum_{p\in \text{ST}} \weight{pq}} \left(\displaystyle\sum_{p\in \text{ST}} \weight{pq}  \left(\matBl \k{pq}  \right) \right) \label{eq.callbacl}\\
0 &= \sum_{\lvec} \left[\matBl \left( \matAa{rq} \k{rq} - \frac{ \left(\displaystyle\sum_{p\in \text{ST}} \weight{pq}  \left(\k{pq}  \right) \right)  }{\displaystyle\sum_{p\in \text{ST}} \weight{pq}} \right)\right] \label{eq.ineq}
\end{align}
where we have used the commutation properties of scalars to factorise $\matBl$. For Eq. \ref{eq.ineq} to hold, the two terms in the brackets must be equal for all values of $\lvec$. $\matAa{rq}$ must therefore be a function of $\lvec$ in addition to the previous parameters. We therefore write:
\begin{align}
\matA{rq} \k{rq} &= \frac{ \left(\displaystyle\sum_{p\in \text{ST}} \weight{pq}  \left(\k{pq}  \right) \right)  }{\displaystyle\sum_{p\in \text{ST}} \weight{pq}}\\
\matA{rq} &= \frac{ \left(\displaystyle\sum_{p\in \text{ST}} \weight{pq}  \frac{\k{pq}}{\k{rq}} \right)  }{ \displaystyle\sum_{p\in \text{ST}} \weight{pq}}\label{eq.resume.0}
\end{align}
We can {simplify} the expression above by decomposing the Fourier kernels into their individual parts:
\begin{align}
\frac{\k{pq}}{\k{rq}} &= \frac{\k{p} \left(\k{q}\right)^{-1} }{\k{p} \left(\k{q}\right)^{-1}}\\
                      &=\k{p} \left(\k{r}\right)^{-1}\\
                      &= \k{pr}
\end{align}
where we have used the property that $\left(\k{p}\right)^{-1}=\left(\k{p}\right)^{*}$. Plugging this back into Eq. \ref{eq.resume.0}, we finally find:
\begin{align}
\matA{rq} &= \displaystyle\sum_{p\in \text{ST}} \k{pr} \frac{\weight{pq} }{\displaystyle\sum_{p'\in \text{ST}} \weight{p'q}} \label{eq.result.decorr1vis}
\end{align}
\pg
This result tells us that the loss of information caused by incoherent beamforming manifests as a baseline-dependent loss factor - that is to say, dependent on the exact baseline formed between two stations at a given time and frequency, not just on the length of this baseline. In other words, in the image plane, incoherent beamforming results in a convolution of the true sky brightness distribution with some position-dependent decoherence PSF. The peak of this position-dependent decoherence PSF will give a measure of how much the measured signal will be affected, and its width will give a measure of how widely the signal is smeared in the sky.

\subsection{Interpreting our Result}

\pg
The decoherence factor calculated for a single superstation visibility, given in Eq. \ref{eq.result.decorr1vis}, depends on a few parameters. Firstly, and most importantly, it is a function of $\lvec$, the angular distance between the position being considered in the sky and the phase centre for the data. Secondly, it is a function of the distance between the superstation's $\uvec{}$ coordinates and those of the stations used to create it. Finally, it is a function of the visibility weighting.

\pg
Let us consider the limits in which the decoherence is negligible, i.e. within which $\matA{rq} \sim 1$. Firstly, if $\lvec=0$ (i.e. when the source is at phase centre), we have:
\begin{align}
\ko{pr} &= 1\\
\matAo{rq} &= \frac{ \displaystyle\sum_{p\in \text{ST}} \weight{pq} }{\displaystyle\sum_{p'\in \text{ST}} \weight{p'q}}\\
          &= 1
\end{align}
and so we are unaffected by this beamforming decoherence, regardless of any other factors.

\pg
Secondly, assuming we are interested in a source located elsewhere than phase centre, let us consider the limit in which we can beamform stations into the superstation at a negligible cost in decoherence. This is particularly relevant for LOFAR, since we could in principle {synthesise} a superstation out of all LOFAR core stations, decreasing the thermal noise in the relevant visibilities by a factor of nearly 7 if phasing up all the core Dutch stations and using the dual mode (where there are 2 HBA sub-stations per station). If our field of view includes only a single source at position $\lvec_0\ne \bm{0}$, and assuming unit weights ($\weight{pq} = const.$), the decoherence factor becomes:
\begin{align}
\matAlo{rq} &= \displaystyle\sum_{p\in \text{ST}} \klo{pr} \
\end{align}

\pg
This quantity obviously tends towards unity as $\uvec{rq}\rightarrow\bm{0}$. This is a trivial limit, however: it simply states that as the difference between the superstation coordinates and its phased antenna coordinates vanishes, so does the decoherence. What interests us, however, is the limit in which the impact of decoherence becomes negligible for a given baseline with an antenna $q$ which is neither part of the antennas being phased up nor the superstation. In other words, what interests us is the limit in which:
\begin{align}
\sum_{\lvec} \left( \matBl \k{rq} \right) &\approx \sum_{\lvec} \left( \matAlo{rq}\matBl \k{rq} \right)
\end{align}
which, with the approximations above, becomes
\begin{align}
\matBlo \klo{rq}  &\approx \displaystyle\sum_{p\in \text{ST}}\klo{pr} \matBlo  \klo{rq}\\
\klo{rq}  &\approx \displaystyle\sum_{p\in \text{ST}}\klo{pr}  \klo{rq}\\
\exp \left(-2\pi i \uvec{rq} \cdot \lvec_0 \right) &\approx
\displaystyle\sum_{p\in \text{ST}}\exp \left(-2\pi i \left( \uvec{rq} + \uvec{pr}\right) \cdot \lvec_0 \right)
\end{align}
which is satisfied when
\begin{align}
\uvec{pr} << \uvec{rq} \forall p \label{eq.limit}
\end{align}
i.e. when the $uvw$ distance between the antennas being beamformed and the resultant superstation is negligible compared to the distance between the superstation $uvw$ coordinates and those of the antennas $q \notin \text{ST}$. This condition is always met by standard VLBI arrays, but LOFAR-VLBI can be an edge case: this limit tells us that, for example, phasing up the Superterp may be fine, but phasing up all core Dutch stations would bring problems (as the distance between the nearest core station to a remote station and the distance between the superterp and said core station could be comparable).

\pg
It should be noted that traditional VLBI satisfies both of these conditions, but large-scale interferometers such as LOFAR and the future SKA may not. If these instruments still wish to make use of VLBI techniques, understanding the exact limits of their applicability will be very useful.

\subsection{Predicting Decoherence for a Full Observation}

\pg
Let us begin by formalising the relationship between visibilities and images made from them. {Because the decoherence factor is a scalar quantity, we will proceed with unpolarised emission from here on out, and therefore reduce our framework to a scalar one. All correlations (and therefore all Stokes images) will experience the effects described from here on out.} A visibility is simply the Fourier transform of the sky brightness distribution sampled at a specific point in Fourier space, which is a function of the $uvw$-coordinates of the antennas forming the baseline. In other words:
\begin{align}
\matV{pq} = \int_{\lvec} \matB_{\lvec} \k{pq}d\lvec
\end{align}
where $\k{pq}$ encodes both the forward Fourier transfer function and the Fourier sampling function. The contribution of this visibility to the position $\lvec$ on a dirty image will then be the inverse Fourier transform of the above. The integral can be thought of as being performed over a {series of ``fringes"} $\dirtyIpq$, each associated to a single visibility:
\begin{align}
\dirtyIpq(\lvec) &= \matV{pq} \weight{pq,t\nu} \left(\k{pq}\right)^*\\
\dirtyI(\lvec)   &= \frac{1}{\int_{pq,t\nu}\weight{pq,t\nu} }\int_{\uvec{pq,t\nu}} \dirtyIpq d\uvec{pq,t\nu}
\end{align}
where $\weight{pq,t\nu}$ is the weight associated to that fringe. Equivalently, this can be written in discrete form as
\begin{align}
\dirtyI(\lvec)   &= \frac{\displaystyle \sum_{pq,t\nu} \dirtyIpq(\lvec)}{\displaystyle \sum_{pq,t\nu} \weight{pq,t\nu}}
\end{align}
\pg By iterating over the cardinal sine coordinates of all the pixels in our image, we can now recreate the so-called ``dirty map" of an observation analytically from the visibilities. 

\pg
Having written the relationship between a set of visibilities for an observation and the resulting dirty image, let us consider the case of an empty sky with a single point source of brightness S at some position $\lvec_0$. This gives us:
\begin{align}
\matB_{\lvec} &= S\delta(\lvec-\lvec_0)\\
\matV{pq} &= \int_{\lvec} \matB_{\lvec} \k{pq}d\lvec\\
          &= S\klo{pq}
\end{align}
and the value of the dirty map at coordinates $\lvec_0$ is then:
\begin{align}
\dirtyIpq(\lvec_0) &= \matV{pq} \weight{pq,t\nu}  \left(\klo{pq}\right)^*\\
                   &= S\klo{pq}\weight{pq,t\nu}  \left(\klo{pq}\right)^*\\
                   &= S\weight{pq,t\nu}  \\
\dirtyI(\lvec_0)     &= \frac{\displaystyle \sum_{pq,t\nu} \dirtyIpq(\lvec_0)}{\displaystyle \sum_{pq,t\nu} \weight{pq,t\nu}}\\
                   &= \frac{S\displaystyle \sum_{pq,t\nu} \weight{pq,t\nu}}{\displaystyle \sum_{pq,t\nu} \weight{pq,t\nu}}\\
                   &= S
\end{align}

\pg
So we see that, by putting our point source through our forward and backward operators, we correctly recover the flux of the source at its known coordinates. We can thus use this formalism, combined with the result of Eq. \ref{eq.result.decorr1vis}, to estimate the peak of the position-dependent decoherence PSF at a given point in the sky.

\pg
For a source with unit brightness at position $\lvec_0$, $B=\delta(\lvec-\lvec_0)$ and the decoherence factor can be written as:
\begin{align}
d_f &= \frac{\dirtyI^{\text{smeared}}(\lvec_0)}{\dirtyI^{\text{unsmeared}}(\lvec_0)}\\
    &=\dirtyI^{\text{smeared}}(\lvec_0)
\end{align}
where the denominator goes away since we have set $S=1$ in this case. To develop further, let us explicitly write an expression for the smeared value of $\dirtyIpq(\lvec_0)$, denoted as $\dirtyIpqsmr(\lvec_0)$:
\begin{align}
\dirtyIpqsmr(\lvec_0) &= \matV{pq} \weight{pq,t\nu} \left(\klo{pq}\right)^*\\
                      &= \left(\int_{\lvec} \matA{pq} \delta(\lvec_0-\lvec)\k{pq}\right) \weight{pq,t\nu}\left(\klo{pq}\right)^*\\
                      &= \matAlo{pq} \weight{pq,t\nu}\klo{pq}\left(\klo{pq}\right)^*\\
                      &= \matAlo{pq}\weight{pq,t\nu}
\end{align}
where, from Eq. \ref{eq.result.decorr1vis} (changing mute indices):
\begin{align}
\matAlo{pq} &= 
\left\{
\begin{array}{@{}ll@{}}
\displaystyle\sum_{r\in \text{ST}} \klo{rp} \frac{\weight{rq} }{\displaystyle\sum_{r\in \text{ST}} \weight{rq}} & \text{if}\ p=p_{ST}\ \text{and}\ q\notin\text{ST} \\
1 & \text{otherwise}
\end{array}\right.
\end{align}
and where $ST$ is the set of antennas being beamformed into superstation antenna $p_{ST}$. Now, to find the proper decoherence factor, we must discard all visibilities with $(p,q)\in\text{ST}$, and use in their stead the visibilities associated with $p_\text{ST}$. Using this method, we can now estimate the suppression factor at any coordinates in the sky as a function of the choice of beamforming stations and the angular distance between those coordinates and phase centre.

%% file: source/sec3.tex
\section{Simulations}\label{sec.3}

\pg
In this section, we will aim to verify the results given in Sec. \ref{sec.2} on a simulated dataset. This dataset is created by slicing 30 minutes of observation from an 8-hour LOFAR HBA observation made in HBA\_DUAL\_INNER mode, resulting in 48 core stations and 14 remote stations. No international stations are present. This allows us to get realistic $uvw$-coordinates and frequency coverage. 

\pg
With this information, we then simulate the visibilities for a sky consisting of a single 1Jy point source at $(l=-0.3,m=0.3)$. We then use NDPPP's StationAdder function \citep{2013A&A...556A...2V} to create a beamformed superstation using all the core stations. This results in a set of visibilities with 63 antennas: the original 62 and one superstation where we expect to see decoherence in the simulated point source. We henceforth refer to this set of visibilities as the averaged visibilities, even though only some of them are affected: this is in contrast to the the control visibilities. These are created by simulating the point source exactly as it ought to be seen for all 63 antennas. There is no noise introduced in this simulation.

\subsection{The simulated dataset}

\begin{figure}[!h]
	\centering
	\includegraphics[width=.9\linewidth]{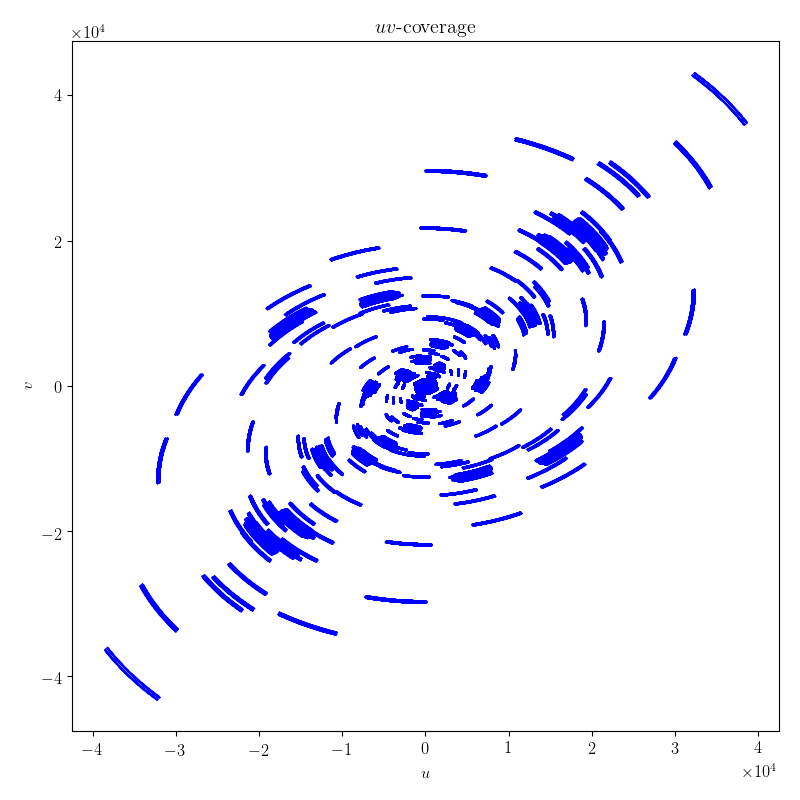} 
	\caption{$uv$-coverage of the dataset chosen to perform our simulations. Quantities are dimensionless.} \label{fig.uvw.coverage}
\end{figure}

\pg
The observation considered was taken on July 28th, 2014, from 1300h to 1400h. We are taking 30 minutes and a single subband for our tests, starting from 13h30. This gives us a bandwidth of 2 kHz centred on 134.86 MHz and 20 channels. The core and remote stations are present, in HBA\_DUAL\_INNER mode \citep[{see}][]{2019A&A...622A...1S}. The $uv$-coverage is shown in Fig. \ref{fig.uvw.coverage}. 

\subsection{Verifying that Beamforming is Correctly Modeled}

\begin{figure*}[t!]
	\centering
	\begin{subfigure}{.49\textwidth}
		\resizebox{\hsize}{!}{\includegraphics[width=.48\linewidth]{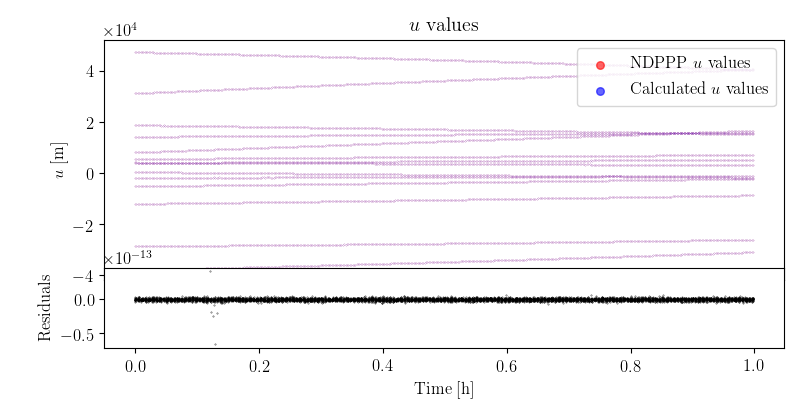}}
		\caption{$u$-coordinates. Residuals of the order
			of $10^{-13}$.} \label{fig.u.resid}
	\end{subfigure}
	\hfill
	\begin{subfigure}{.49\textwidth}
	\resizebox{\hsize}{!}{\includegraphics[width=.48\linewidth]{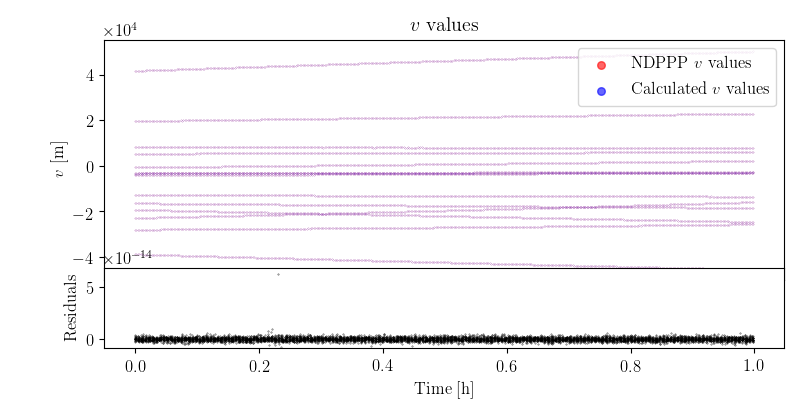}}
	\caption{$v$-coordinates. Residuals of the order
	of $10^{-14}$.} \label{fig.v.resid}
\end{subfigure}
	\hfill
	\begin{subfigure}{.49\textwidth}
	\resizebox{\hsize}{!}{\includegraphics[width=.48\linewidth]{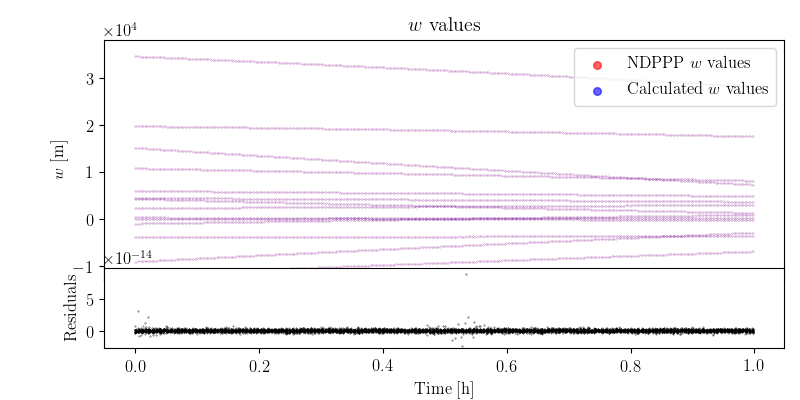}}
	\caption{$w$-coordinates. Residuals of the order
		of $10^{-14}$.} \label{fig.w.resid}
	\end{subfigure}
	\hfill
	\begin{subfigure}{.49\textwidth}
	\resizebox{\hsize}{!}{\includegraphics[width=.48\linewidth]{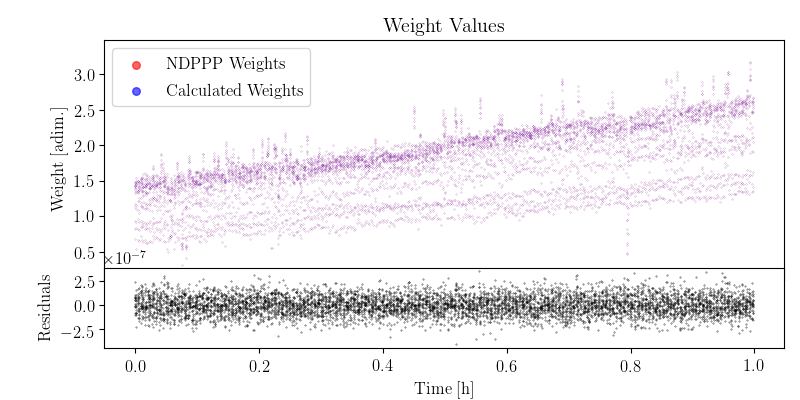}}
	\caption{Weights. Residuals of the order
		of $10^{-14}$.} \label{fig.weights.resid}
	\end{subfigure}	
	\hfill
	\begin{subfigure}{.49\textwidth}
		\resizebox{\hsize}{!}{\includegraphics[width=.48\linewidth]{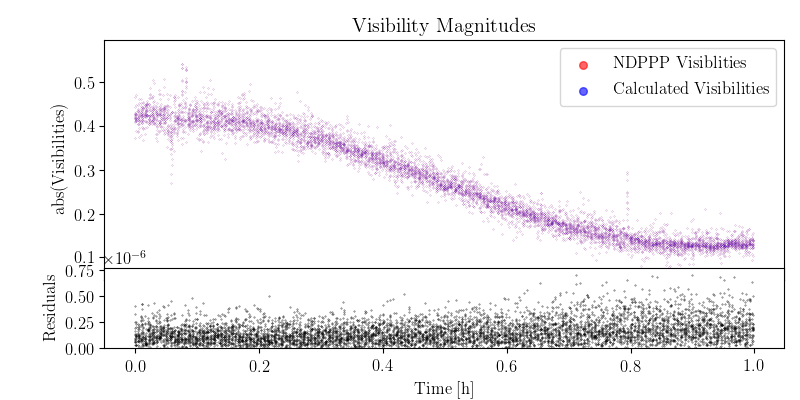}} 
		\caption{Visibility amplitudes. Residuals of the order of $10^{-7}$.} \label{fig.vis.resid.amp}
	\end{subfigure}
	\hfill
	\begin{subfigure}{.49\textwidth}
		\resizebox{\hsize}{!}{\includegraphics[width=.48\linewidth]{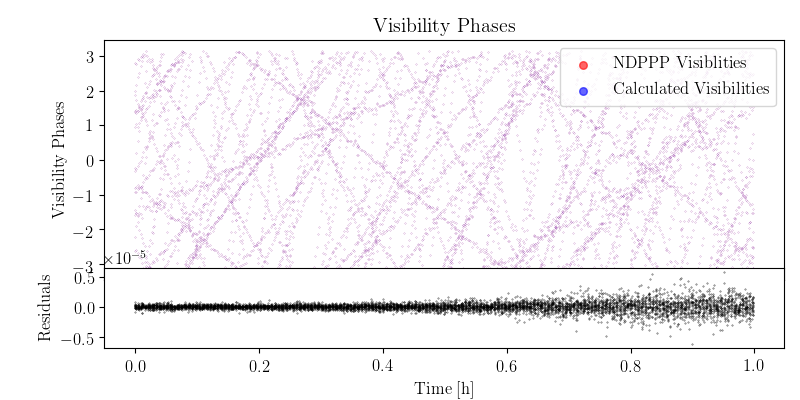}}
		\caption{Visibility phases. Residuals of the order of $10^{-6}$.} \label{fig.vis.resid.phase}
	\end{subfigure}
	\caption{\label{fig.resids} Values for various quantities for baselines pointing towards the beamformed superstation, calculated with
		NDPPP and by the authors. Residuals under the plot: these are relative residuals, meaning that they have been 
		normalised by the associated quantity value.}
\end{figure*}

\pg
Here, our aim is to show that the equations given in Sec. \ref{sec.2} - Eq. \ref{eq.STuvw}, Eq. \ref{eq.weights}, and Eq. \ref{eq.vis.st} - correctly model the behaviour of the LOFAR post-processing beamforming software - specifically, NDPPP's StationAdder function. To this end, we take the weights, $uvw$-coordinates and control visibilities for the beamformed antennas pointing to individual remote antennas at a given time, apply  Eqs. \ref{eq.STuvw}, \ref{eq.weights} and \ref{eq.vis.st}, and compare these visibilities with those of the averaged visibilties for that baseline. The values of the residual $uvw$-coordinates are given in Figures \ref{fig.u.resid}, \ref{fig.v.resid}, and	 \ref{fig.w.resid}, while the weights are shown in Fig. \ref{fig.weights.resid} and the visibility phase and amplitude in Figs. \ref{fig.vis.resid.amp} and \ref{fig.vis.resid.phase}, respectively. These are all relative residuals, meaning that they are normalised by the measured values. In other words, if the residual $x_r$ between a value $x$ and its measure $x_m$ is $x_r = x-x_m$, we are plotting $\tilde{x}=\frac{x-x_m}{x_m}$. : This removes patterns in the residuals which are due to varying amplitudes in the values being computed, rather than to systemic errors.

\pg
As these residual figures show, we calculate the same values as NDPPP for all quantities of interest, up to machine noise (i.e. up to the precision of the averaging function of either python or NDPPP). The loss of coherence outlined in Sec. \ref{sec.2} (and starkly visible in Fig. \ref{fig.vis.resid.amp}) therefore applies to any sources away from phase centre in observations which use NDPPP's StationAdder routine to form a superstation. Furthermore, the simulations in this section accurately depict this loss of coherence.

\subsection{Predicting Impact of Decoherence on Simulated Data}

\pg
The visibilities averaged in the previous section are the simulated visibilities corresponding to a single 1 Jy point source away from phase centre. The amplitude of these visibilities would be expected to be unity for all measurements. We can immediately see the impact of simple averaging in post-processing superstation beamforming visibilities by inspecting Fig. \ref{fig.vis.resid.amp}: whereas we expect the visibility amplitude to be 1 at all points of measurement, we instead find constant suppression. 

%

\pg
However, althought this decoherence is immediately obvious in visibility space, its impact in image space is of greater interest. To characterise it, we have simulated a single point source at increasing distances from phase centre. We have then created two dirty images from each simulated set of visibilities. One includes the beamformed superstation formed using all LOFAR core stations (and therefore affected by decoherence), flagging all core stations during imaging. The other does not include the beamformed superstation, but uses all the core stations. Both are then effectively two ``similar" images of the source, but one is affected by smearing.

\begin{figure*}[t!]
	\centering
	\begin{subfigure}{.43\textwidth}
		\resizebox{\hsize}{!}{\includegraphics{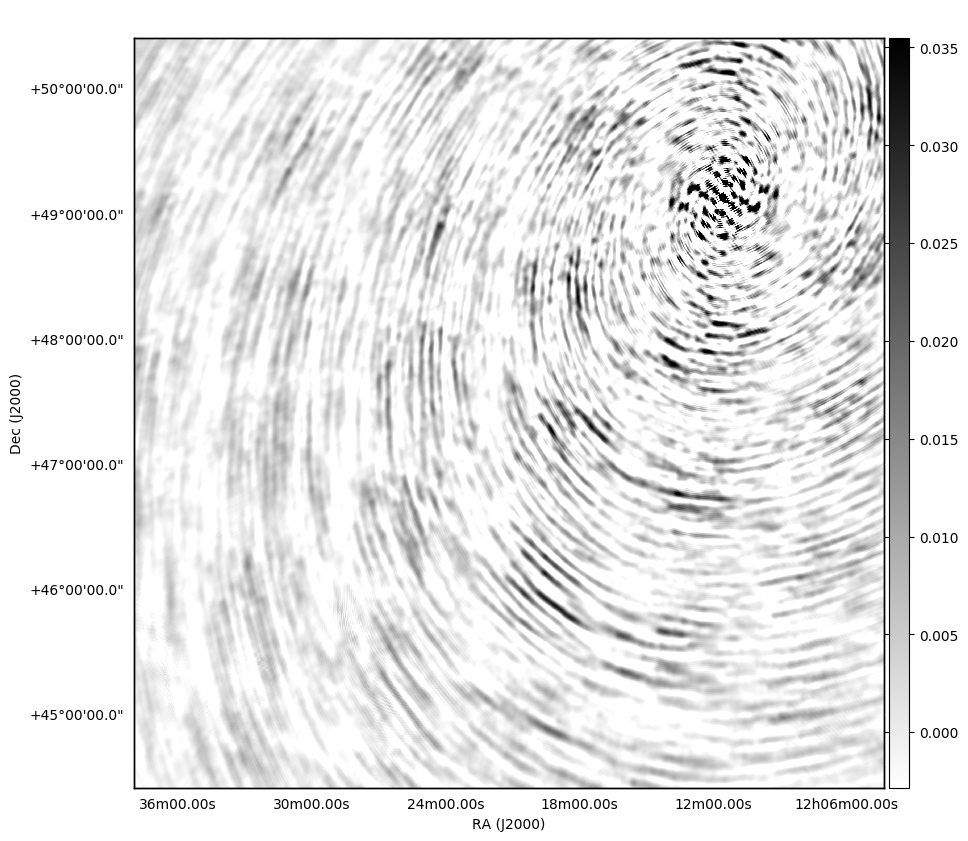}}
		\caption{\label{image.simu.nosmear.noST} Dirty map of our simulated point source, made from the control visibilities (no averaging) and without using the superstation. No noise present. Peak flux value is 1.77 Jy.}
	\end{subfigure}
	\hfill
	\begin{subfigure}{.43\textwidth}
		\resizebox{\hsize}{!}{\includegraphics{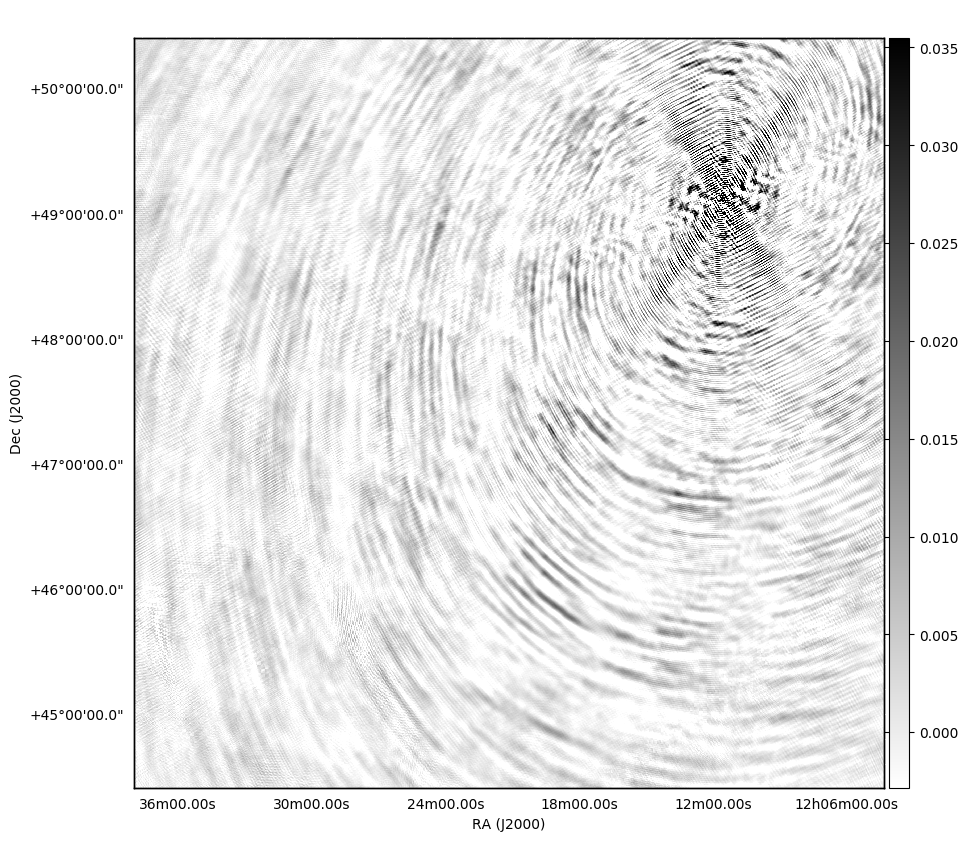}}
		\caption{\label{image.simu.nosmear.noCS} Dirty map of our simulated point source, made from the control visibilities (no averaging) and without using the core stations, using the superstation instead. No noise present. Peak flux value is 1.77 Jy.}
	\end{subfigure}
	\hfill
	\begin{subfigure}{.43\textwidth}
		\resizebox{\hsize}{!}{\includegraphics{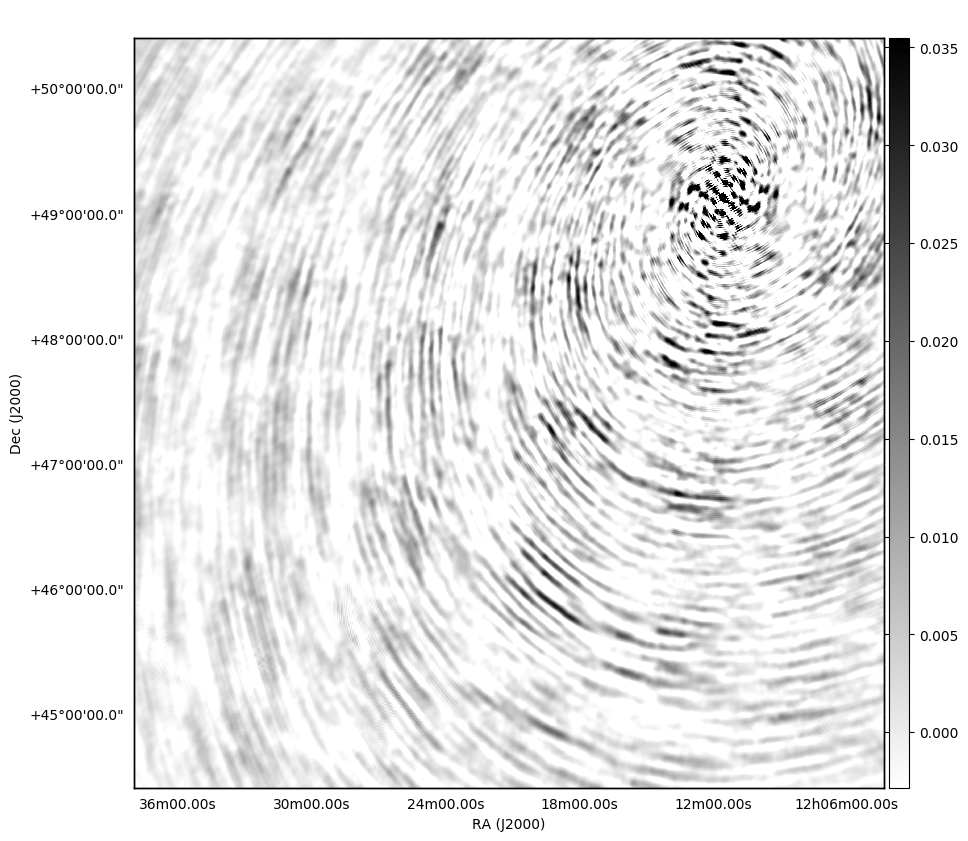}}
		\caption{\label{image.simu.smeared.noST} Dirty map of our simulated point source, made from the averaged visibilities (using NDPPP StationAdder) and without using the superstation. No noise present. Peak flux value is 1.74 Jy.}
	\end{subfigure}
	\hfill
	\begin{subfigure}{.43\textwidth}
		\resizebox{\hsize}{!}{\includegraphics{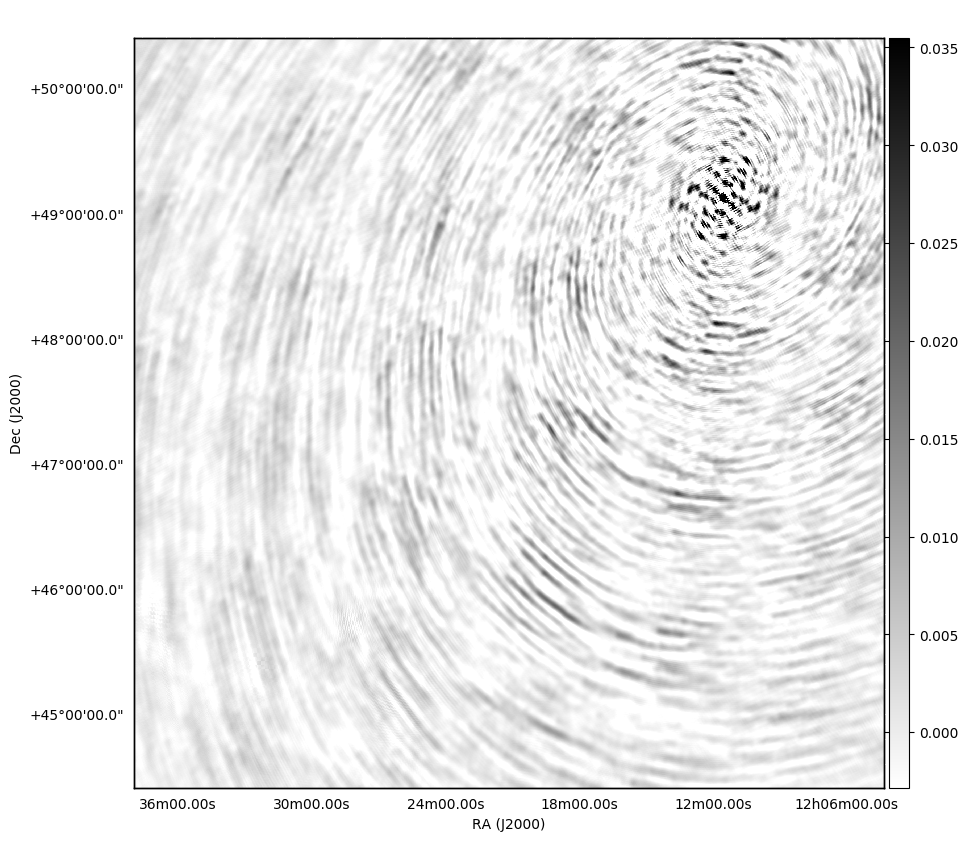}}
		\caption{\label{image.simu.smeared.noCS} Dirty map of our simulated point source, made from the averaged visibilities (using NDPPP StationAdder) and without using the core stations, using the superstation instead. No noise present. Peak flux value is 1.09 Jy.}
    \end{subfigure}
	\caption{\label{image.simudecorr} Dirty maps of a single point source simulated into the visibilities. The images on the left are made using the simulated visibilities directly. The images on the right are made using the superstation visibilities: at the top they are simulated directly, and at the bottom they result from post-processing beamforming. We see that the higher spatial frequencies are suppressed: this accounts for the absence of the ``x"-shaped ``artefact" in Fig. \ref{image.simu.smeared.noCS}: this artefact is in fact the part of the PSF which corresponds to our high spatial frequencies.}
\end{figure*}

\pg
Let us begin by verifying that our simulations show the predicted behaviour  when using NDPPP's post-processing beamforming routine on our simulated visibilities, but not when simulating the visibilities directly into the superstation.  This is shown in Fig. \ref{image.simudecorr}. We see that the dirty maps made without the superstation are equivalent, but that the dirty maps made using the superstation differ, with the small spatial scales suppressed when using the post-processing beamforming. This accounts for the presence of a cross-shaped ``artefact" in Fig. \ref{image.simu.nosmear.noCS}, which disappears in Fig. \ref{image.simu.smeared.noCS}: this ``artefact" is in fact the high spatial frequencies of the PSF, precisely that which is supressed by our post-processing averaging. Although it is hard to see, this results in the peak flux in Fig. \ref{image.simu.smeared.noCS} being reduced to only $57\%$ of what it is in all other images.

\pg
We make one such set of 4 images for a range of $l$-values, keeping $m$ zero. For each such set of images, we find the pixel with the highest flux value in the dirty map made using the core stations. For this pixel, we then find the flux value in the image made using the superstation. This is the measured decorrelation factor, $d_f$.

\pg
We then create an inverse Fourier kernel for the associate $(l,m)$-values, and compute the value expected in both cases at the exact position of the source. This is done analytically, without use of any imager package.

\begin{figure}[!h]
	\centering
	\includegraphics[width=.99\linewidth]{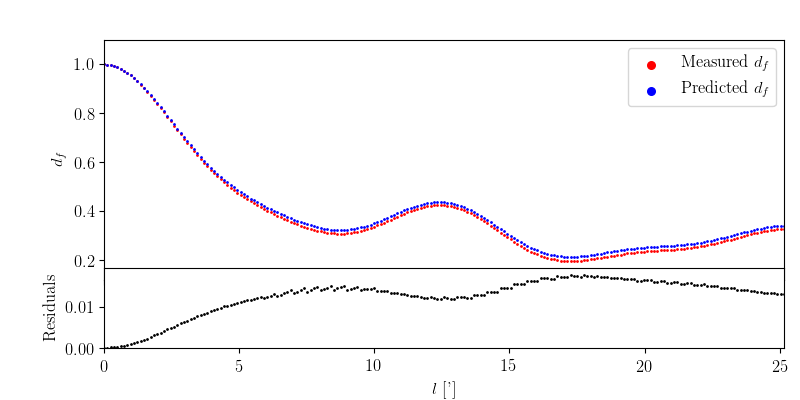} 
	\caption{\label{fig.ldecorr} Decorrelation function as a function of distance from phase centre, for a point source located at that position and with a given set of antennas chosen to form a superstation. Here, the residuals are absolute residuals, not relative residuals. This is to show the quantisation in the residuals. $l$ is in units of arcminutes.} 
\end{figure}

\pg
Finally, we plot both of those decorrelation factors as functions of $l$ in Fig. \ref{fig.ldecorr}. As we can see, both curves are in agreement, to within a few percent. The source of the disparity is very likely found in the quantisation of the Fourier kernel necessary for imaging packages: the imager does not calculate the values of the sky brightness distribution at every $l,m$ value, but only at those on its grid. This is likely why, after about $12'$, the residuals tend to have the same values for every $\sim 4$ values of $l$. 

\pg
Note that decorrelation becomes very strong very fast: this is because we have chosen to use every single core LOFAR station for our beamformed superstation, which exaggerates the strength of the decorrelation. 
LOFAR observations which use only the Superterp (usually formed with the 6 innermost core stations) will be much less affected by this behaviour. However, it has become standard practice in LOFAR-VLBI to use all core stations to form a superstation, still referred to as the Superterp. As such, in the context of LOFAR-VLBI at the time of writing, this decorrelation plot is representative of the signal loss one could expect in real observations which do not use the international stations. This signal loss has yet to be characterised, and will depend on the international stations being used, but - assuming that signal can be calibrated and recovered on those international baselines - they are expected to suffer less decorrelation as they are closer to the limit of negligibility given in Eq. \ref{eq.limit}.

\pg
The structure in the residuals is likely due to sky coordinates being quantised in images; the overall error never rises above 1\% of the prediction, however, and is therefore neglibible. The bump around $l\sim 12'$ is discussed further in Sec. \ref{sec.conclusion}, ``Correcting Decoherence".

\pg
In conclusion, Fig. \ref{fig.ldecorr} conclusively shows that the decoherence introduced by post-processing beamforming is not only analytically understood, but accurately modeled by our predictions.


%% file: source/sec4.tex

\section{Application to Real Data}\label{sec.4}

\pg
In this section, we take a single subband of the full 8 hours of observations from which a slice was taken to create the simulations shown in Sec. \ref{sec.3}. This data is then calibrated and imaged with certain constraints, in order to confirm that the behaviour described in Sec. \ref{sec.3} applies to real data as well. We will begin with a description of the observation itself, and then explain the choices made to confirm the presence of source suppression. This will entail explaining our calibration and imaging procedures, along with certain flagging choices made to ensure we compare like to like.

\subsection{Description of Observation \& Data Reduction}

\pg
The dataset used here was a single subband ($\nu_0=128.3188$ MHz, $\Delta\nu=195.3$ kHz, split into 8 channels of $24.4$ kHz each) of an 8-hour LOFAR HBA observation of the Extended Groth Strip, pointing at ($\alpha,\delta$)=(14:17:00.00,6+52.30.00.00) and taken on 28/9/2014 from noon to 8pm, UTC. The observation was done in HBA\_DUAL\_INNER configuration, where the core stations (which are usually made up of 48 beamformed antennas) are ``split" into two phased arrays with 24 antennas each. The visibilities were averaged to 1 measurement per 2 seconds per baseline.

\pg
We began by calibrating the dataset using killMS \citep{2015MNRAS.449.2668S}. One calibration solution was found per 4 channels and per 8 seconds. Calibration was done using the best high-resolution model of 3C295 currently available at LOFAR frequencies (Bonnassieux et al., in prep.). Once this was done, this dataset was used to create two new ones using LOFAR's New Default Pre-Processing Pipeline \citep[NDPPP: cf. ][]{2013A&A...556A...2V}. Specifically, one dataset was created by reading the calibrated visibilities and writing them into a new dataset, while forming a superstation from all the core stations in the original dataset and flagging out the international stations (along with two remote stations which were found to have poor-quality calibration solutions during calibration, RS210 and RS509). The other dataset was formed by keeping only core-remote, remote-core and remote-remote baselines (excluding, once again, RS210 and RS509).

\pg
The international stations were removed because the model used to calibrate is not yet good enough to ensure that their calibration solutions are correct, as they resolve the brightest sources out. They thus lead us outside of the scope of our formalism. By flagging them, we ensure that the calibrator can be considered as a point source, and thus that our calibration solutions are reliable.
The removal of core-core baselines was done because autocorrelations are not preserved for LOFAR imaging. Thus, when forming the superstation from all core stations, all core-core baselines get removed - since all core stations are now ``one" station, every baseline between different core stations is treated as an autocorrelation, and therefore discarded. As a consequence, flagging the core-core baselines is necessary to ensure that the comparison between our datasets is valid. Without it, the first dataset has many, many more visibilities - and the observation therefore has a different sensitivity, and the comparison becomes invalid, since it is the signal loss due specifically to decoherent superstation formation that is of interest to us here. Our final dataset therefore includes no core-core baselines at all.

\subsection{Source Extinction}

\begin{figure*}[t!]
	\centering
	\begin{subfigure}{.43\textwidth}
		\resizebox{\hsize}{!}{\includegraphics{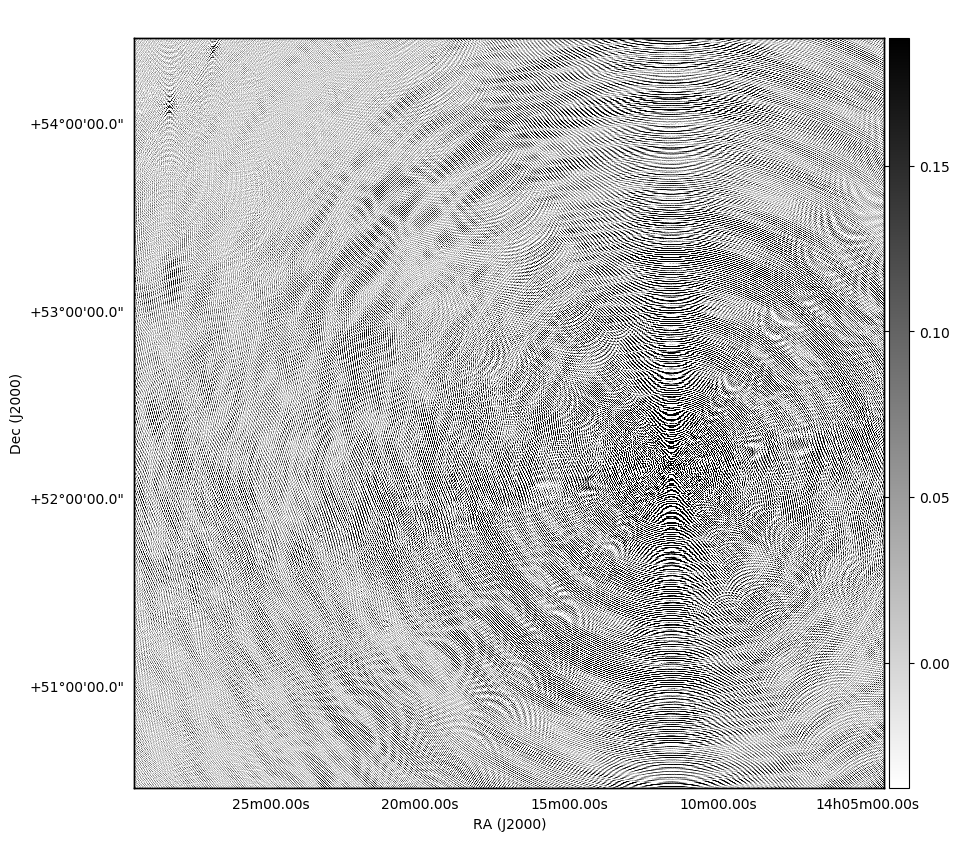}}
		\caption{\label{image.real.nost.dirty} Dirty map of the Extended Groth Strip and 3C295, made using the core and remote stations while flagging the core-core baselines. This image shows apparent flux, not intrinsic flux.}
	\end{subfigure}
	\hfill
	\begin{subfigure}{.43\textwidth}
		\resizebox{\hsize}{!}{\includegraphics{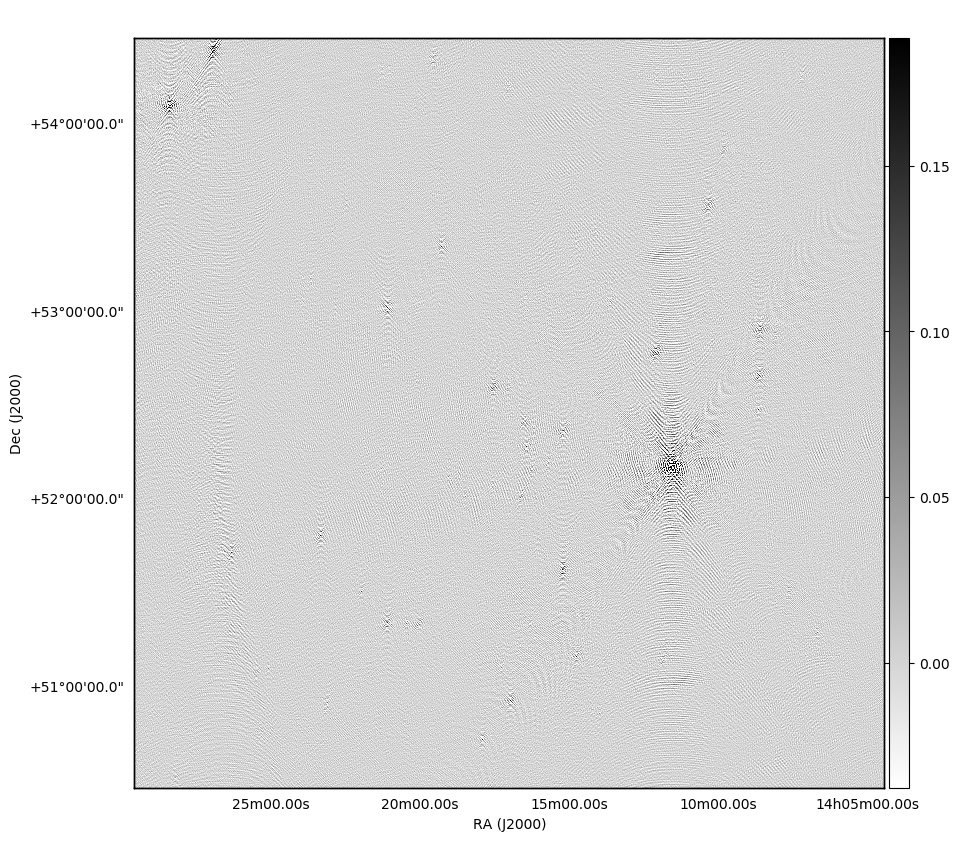}}
		\caption{\label{image.real.nost.restored} Restored map of the Extended Groth Strip and 3C295, made using the core and remote stations while flagging the core-core baselines. This image shows apparent flux, not intrinsic flux.}
	\end{subfigure}
	\hfill
	\begin{subfigure}{.43\textwidth}
		\resizebox{\hsize}{!}{\includegraphics{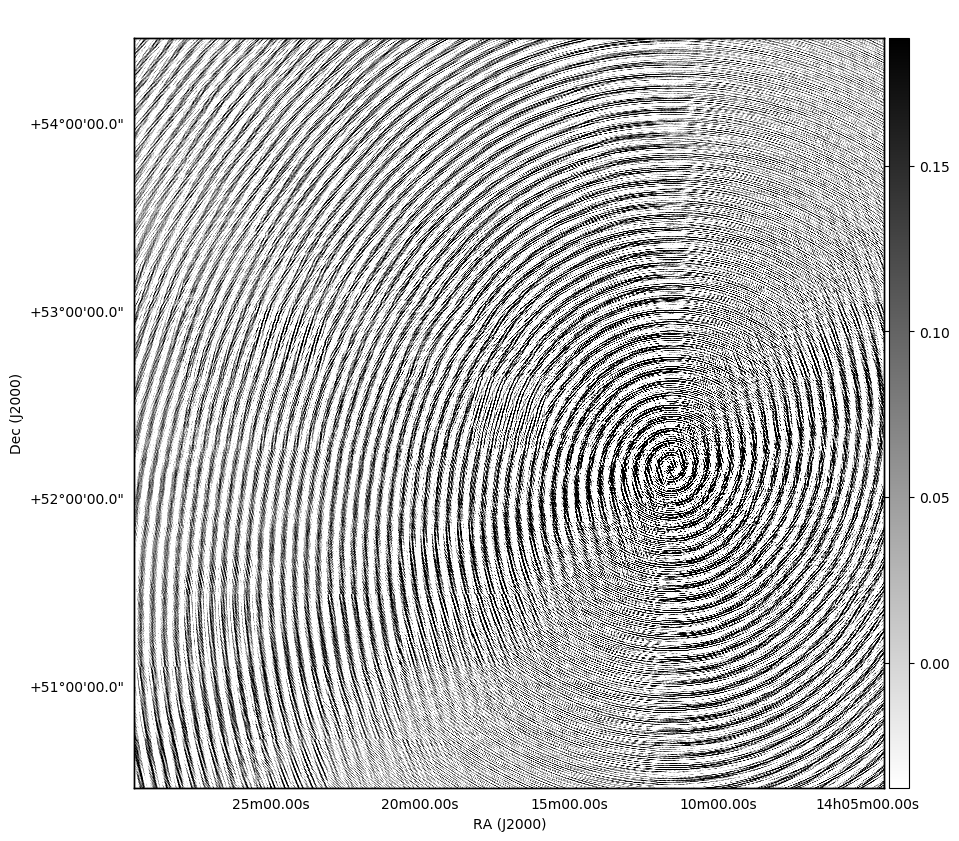}}
		\caption{\label{image.real.nocs.dirty} Dirty map of the Extended Groth Strip and 3C295, made using the remote stations and a superstation from the core stations. This image shows apparent flux, not intrinsic flux.}
	\end{subfigure}
	\hfill
	\begin{subfigure}{.43\textwidth}
		\resizebox{\hsize}{!}{\includegraphics{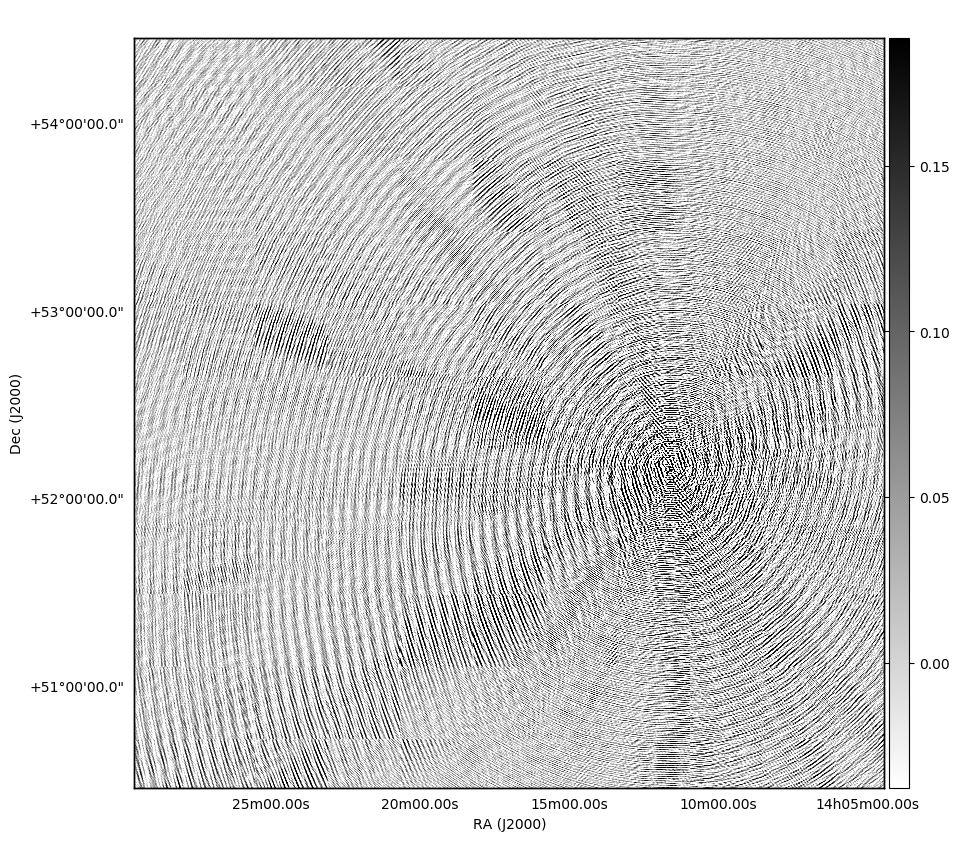}}
		\caption{\label{image.real.nocs.restored} Restored map of the Extended Groth Strip and 3C295, made using the remote stations and a superstation from the core stations. This image shows apparent flux, not intrinsic flux.}
	\end{subfigure}
	\caption{\label{image.realdecorr} Dirty \& restored maps of the sky showing the impact of incoherent post-processing superstation beamforming. Colour scales and pixel coordinates are matched in all images, and all units are Jy/bm. 3C295 is the very prominent, very bright source which dominates the field. While the PSF changes as a result (which is an expected and normal outcome), signal is also lost: this can be seen most clearly in the two North-Eastern sources in Figs. \ref{image.real.nost.dirty},\ref{image.real.nost.restored} vanishing from Figs. \ref{image.real.nocs.dirty},\ref{image.real.nocs.restored}.}
\end{figure*}

\pg
Each dataset was imaged using the same imaging parameters: a cell size of $1"$, 15k$\times$15k pixels, an inner $uv$-cut of 10km (thereby excluding the shortest baselines to ensure consistency between our calibration and imaging) and Briggs weighting with a robust parameter value of 2 \citep{1995AAS...18711202B}. The only differences between the imaging runs were the output names and dataset used. The dirty maps (Figs. \ref{image.real.nost.dirty} and \ref{image.real.nocs.dirty}) are simply the inverse Fourier transform of the visibilities, and therefore map the sky brightness distribution convolved with the instrument response, or PSF. The restored maps (Figs. \ref{image.real.nost.restored} and \ref{image.real.nocs.restored}) are the result of running the dirty maps through a deconvolution algorithm, stopped in both cases by running out of major iterations (20). As such, the most prominent sidelobes of the brightest sources are reduced, allowing fainter sources to appear. 

\pg
A few things in Fig. \ref{image.realdecorr} warrant comment. Firstly, the difference in ``source morphology" between Fig. \ref{image.real.nost.dirty} and Fig. \ref{image.real.nocs.dirty} is due to the different $uv$-coverage in both observations, leading to different PSFs. This is normal and expected.

\pg
Secondly, we can see that the deconvolution did not give equivalent results in both cases. While some artefacts remain around 3C295 (which is the brightest source in the field by far) in Fig. \ref{image.real.nost.restored}, the shift from  Fig. \ref{image.real.nost.dirty} to  Fig. \ref{image.real.nost.restored} is what one expects from comparing dirty and restored maps: sources in the field which were previously hidden in the sidelobes become apparent. The shift from Fig. \ref{image.real.nocs.dirty} to  Fig. \ref{image.real.nocs.restored}, however, gives no such improvement. Indeed, the deconvolution seems to deteriorate the image rather than improve it. This is in fact expected. Indeed, the signal loss introduced by decoherent superstation beamforming does not manifest simply as a flux loss: indeed, it manifests as a \textit{baseline-dependent} extinction factor. In other words, it manifests as a modulation of the PSF, which the deconvolution algorithm does not take into account. It then attempts to deconvolve with an incorrect PSF, resulting in overall deterioration. This is particularly obvious in this image due to 3C295's massive brightness, but is expected to occur - less obviously but no less problematically - for all sources in the field. This deconvolution issue likely accounts for the aliasing effect visible in Fig. \ref{image.real.nocs.dirty}. As a consequence of this, a fair comparison between the images ought to be done using the dirty maps rather than the restored maps. 

\pg
Thirdly, then, we can note one very interesting fact by visual inspection of Fig. \ref{image.real.nost.dirty} and Fig. \ref{image.real.nocs.dirty}. In the former, two sources are present in the northeast corner (top left). They are absent in the second. These two sources also show up in Fig. \ref{image.real.nost.restored}, so they are not just some kind of strange resonance: they are physical sources. In other words, we have firsthand, visible evidence that superstation formation results in source suppression, all other things being equal. By taking the ratio of the highest pixel values in both dirty images, we calculate a measured decoherence factor for 3C295: $d_{f,m}=20.77\%$. In other words, the brightest pixel in the image made using the superstation is only about $20\%$ of that in the image using the core stations (without the core-core baselines).

%% file: source/conclusion.tex

\section{Conclusion \& Future Work}\label{sec.conclusion}

\pg
Thus far, this paper has focused on a demonstration of the effects of an interferometric technique. Ultimately, however, the purpose of techniques is to be used, and used wisely. We therefore wish to provide scientists with guidelines to help them make the most out of this technique while mitigating, or at least recognising, potential negative consequences on their scientific analysis. 


\pg
As such, our conclusion consists of two discussions. The first aims to help recognise and avoid negative effects from superstation formation, providing practical advice for scientists who are literate in radio interferometry but not specialised in its techniques. The second is more appropriate to an audience of specialists who may wish to take this effect into account in their pipelines or data reduction software, and will discuss possible paths forward to do so.
%
%

\subsection{Avoiding decoherence}

\pg
The first, and most obvious, conclusion of this paper is also the most important: if you seek to improve your SNR on long baselines, where signal is scarcer, by performing post-processing beamforming, you will systematically lose signal if using currently-existing data reduction suites. This signal loss will be greater as distance from the phase centre increases. Accordingly, should this be the approach chosen to reduce an interferometric dataset, care must be taken to rephase the observation on to the target source before forming the superstation. 

\pg
This effect will be negligible if the distance between the stations being beamformed is negligible compared to the distance between these stations and others in array, as expressed in \ref{eq.limit}. For a single baseline, the exact expression is given in Eq. \ref{eq.result.decorr1vis}. It is also negligible near the phase centre at which the beamforming was made. As such, this effect is expected to be negligible for ``standard" VLBI, which will usually only be concerned with a single target at phase centre, and where the distances between the stations being beamformed into a superstation will be neglibible compared to the distance between these stations and more distant stations. The LOFAR-VLBI pipeline is also largely unaffected, as it takes care to rephase around each source, average, combine stations, and then calibrate. 

\pg
If using a ``naive" approach to create a widefield image with a beamformed superstation, the impact of decoherence will manifest as artefacts around sources (caused by the deconvolution of a source with the wrong, ``unsmeared" PSF) and, for point sources, in a net loss in integrated flux. This can be handled in the same way as the beam in apparent-flux images: by predicting the expected decoherence at the positions of the sources in a field, the ``true" flux may be recovered. For LOFAR, assuming that you choose to {use} all core stations to form your superstation, and that your observation includes all international stations present in 2015, this will result in the decoherence factors as a function of distance from phase centre given in Fig. \ref{fig.ldecorr}.

\pg
As such, for ``standard" widefield observations with international stations, it is \textit{not} recommended to form up the Superterp in this way at the time of writing if any emission from the source of interest lies farther than $1'$ away from phase centre (either due to the size of the source or its location relative to phase centre). This will ensure that signal suppression remains below $\sim5\%$. Once the methods described in the section below are implemented, it may become safe to form the Superterp from core stations. The best way to ensure that one's scientific analysis is not negatively affected remains to perform the same test as shown here: image the same dataset with and without superstation beamforming. This can be done with a small subset of the overall dataset without negatively impacting the test.

\pg
Finally, in the specific case of LOFAR-VLBI in the LBA regime (i.e. at the lowest frequencies), this problem of superstation beamforming will likely be done away with in the near future thanks to NenuFAR \citep[New Extension in Nancay Upgrading LOFAR - cf][ and Zarka et al., The Low Frequency Radiotelescope NenuFAR, Exp. Ast., in preparation]{2012sf2a.conf..687Z,icatt2015}. NenuFAR is a compact array of 19 tiles of 96 LWA-like antennas \citep{2012PASP..124.1090H}. These are much more sensitive than LOFAR LBA antennas out of a narrow spectral range centered at 58 MHz, the peak response of LBA antennas, and are laid out within a diameter of 400m. {It can operate in several modes, 
notably as a LOFAR superstation created by connecting the 96 tiles of 19 antennas to the LOFAR-FR606 receivers.}
The beamforming of the entire LOFAR core prior to correlation with LOFAR stations should provide a superstation with $>$19 times the sensitivity of a standard international LOFAR station, without any of the drawbacks described in the present paper. It could thus naturally serve all the purposes that a superstation can (``anchoring" the calibration solutions for the longest baselines) without the current superstation formation cost of losing the core-core visibilities when solving for international station gains (which can also serve to ``anchor" the core station gains, and thus contribute to anchoring the international station gains in turn). {However, being located on the outskirts of the array, it would not be a good substitute to the core superstation, but should provide a fantastic additional anchor. As such, NenuFAR can be expected to be very beneficial to LOFAR-VLBI observations at the lowest frequencies, once it becomes operational as an alternative to the French LOFAR station.}

\subsection{Correcting decoherence}

\pg
Because decoherence due to post-processing beamforming can be predicted, it can in principle be modeled away. The most obvious way to do this is the method proposed above: simply predict the extent of coherence lost at the position of the sources in the sky model used to calibrate, and apply the appropriate decorrelation correction factor at each position. If the sky model consists of unresolved point sources, this is an equivalent approach to applying the interferometric array beam response to the model, and is expected to give equivalent results in correcting for its associated effect. Creating such a script will be the subject of future work. The main issue here is that the decoherence will depend on the dataset: its $uv$-coverage, the choice of stations to beamform, etc will all have an impact. This script is therefore not entirely trivial to create in a user-friendly way.

\pg
Furthermore, in the presence of diffuse emission, this approach will not be sufficient. We have shown that the decoherence not only introduced some decoherence factor, but also changes the local PSF. This is very significant: it means that decoherence will necessarily introduce deconvolution artefacts around sources (since the deconvolution will be performed with the wrong PSF). This will have much greater consequences for diffuse emission, as these artefacts will accumulate and further bias the deconvolved map. As such, an ideal solution would be to perform much the same operation as DDFacet \citep{2018A&A...611A..87T} uses with baseline-dependent averaging to model away time/frequency smearing: apply the decoherence directly to the PSF, and deconvolve with the smeared PSF. Making this an option in DDFacet is part of our expected future work.

\pg
In continuation with this notion of baseline-dependent averaging, we can propose one final solution to this issue, which would be the ideal one. Post-processing beamforming introduces decoherence because it uses, effectively, a tophat function for its averaging. This explains the odd ``bump" seen in Fig. \ref{fig.ldecorr}: compare it with smearing functions shown in \cite{2018MNRAS.477.4511A}. It is simply the image-space consequence of using a tophat averaging function in Fourier-space: it introduces a sinc smearing in image-space. Therefore, if one were to use baseline-dependent averaging during superstation formation, as explained in \cite{2018MNRAS.477.4511A}, this issue could be done away with by simply specifying a desired Field of View (which the cited paper refers to as ``FoV shaping"), which will determine the period of the baseline-dependent averaging sinc function used to create the visibilities during superstation formation. Implementing this application of FoV shaping in existing software is beyond the scope of this paper, and proving both that this method corrects for the problem and implementing it in existing astronomy software will be the subject of future work.

%% file: paper.bbl
\begin{thebibliography}{18}
\expandafter\ifx\csname natexlab\endcsname\relax\def\natexlab#1{#1}\fi

\bibitem[{{Atemkeng} {et~al.}(2018){Atemkeng}, {Smirnov}, {Tasse}, {Foster},
  {Keimpema}, {Paragi}, \& {Jonas}}]{2018MNRAS.477.4511A}
{Atemkeng}, M., {Smirnov}, O., {Tasse}, C., {et~al.} 2018, \mnras, 477, 4511

\bibitem[{{Briggs}(1995)}]{1995AAS...18711202B}
{Briggs}, D.~S. 1995, in Bulletin of the American Astronomical Society,
  Vol.~27, American Astronomical Society Meeting Abstracts, 1444

\bibitem[{{Hicks} {et~al.}(2012){Hicks}, {Paravastu-Dalal}, {Stewart},
  {Erickson}, {Ray}, {Kassim}, {Burns}, {Clarke}, {Schmitt}, {Craig},
  {Hartman}, \& {Weiler}}]{2012PASP..124.1090H}
{Hicks}, B.~C., {Paravastu-Dalal}, N., {Stewart}, K.~P., {et~al.} 2012, \pasp,
  124, 1090

\bibitem[{{Jackson} {et~al.}(2016){Jackson}, {Tagore}, {Deller}, {Mold{\'o}n},
  {Varenius}, {Morabito}, {Wucknitz}, {Carozzi}, {Conway}, {Drabent},
  {Kapinska}, {Orr{\`u}}, {Brentjens}, {Blaauw}, {Kuper}, {Sluman}, {Schaap},
  {Vermaas}, {Iacobelli}, {Cerrigone}, {Shulevski}, {ter Veen}, {Fallows},
  {Pizzo}, {Sipior}, {Anderson}, {Avruch}, {Bell}, {van Bemmel}, {Bentum},
  {Best}, {Bonafede}, {Breitling}, {Broderick}, {Brouw}, {Br{\"u}ggen},
  {Ciardi}, {Corstanje}, {de Gasperin}, {de Geus}, {Eisl{\"o}ffel}, {Engels},
  {Falcke}, {Garrett}, {Grie{\ss}meier}, {Gunst}, {van Haarlem}, {Heald},
  {Hoeft}, {H{\"o}randel}, {Horneffer}, {Intema}, {Juette}, {Kuniyoshi}, {van
  Leeuwen}, {Loose}, {Maat}, {McFadden}, {McKay-Bukowski}, {McKean}, {Mulcahy},
  {Munk}, {Pandey-Pommier}, {Polatidis}, {Reich}, {R{\"o}ttgering},
  {Rowlinson}, {Scaife}, {Schwarz}, {Steinmetz}, {Swinbank}, {Thoudam},
  {Toribio}, {Vermeulen}, {Vocks}, {van Weeren}, {Wise}, {Yatawatta}, \&
  {Zarka}}]{2016A&A...595A..86J}
{Jackson}, N., {Tagore}, A., {Deller}, A., {et~al.} 2016, \aap, 595, A86

\bibitem[{{Kappes} {et~al.}(2019){Kappes}, {Perucho}, {Kadler}, {Burd},
  {Vega-Garc{\'\i}a}, \& {Br{\"u}ggen}}]{2019A&A...631A..49K}
{Kappes}, A., {Perucho}, M., {Kadler}, M., {et~al.} 2019, \aap, 631, A49

\bibitem[{{Morabito} {et~al.}(2016){Morabito}, {Deller}, {R{\"o}ttgering},
  {Miley}, {Varenius}, {Shimwell}, {Mold{\'o}n}, {Jackson}, {Morganti}, {van
  Weeren}, \& {Oonk}}]{2016MNRAS.461.2676M}
{Morabito}, L.~K., {Deller}, A.~T., {R{\"o}ttgering}, H., {et~al.} 2016,
  \mnras, 461, 2676

\bibitem[{{Shimwell} {et~al.}(2019){Shimwell}, {Tasse}, {Hardcastle}, {Mechev},
  {Williams}, {Best}, {R{\"o}ttgering}, {Callingham}, {Dijkema}, {de Gasperin},
  {Hoang}, {Hugo}, {Mirmont}, {Oonk}, {Prandoni}, {Rafferty}, {Sabater},
  {Smirnov}, {van Weeren}, {White}, {Atemkeng}, {Bester}, {Bonnassieux},
  {Br{\"u}ggen}, {Brunetti}, {Chy{\.z}y}, {Cochrane}, {Conway}, {Croston},
  {Danezi}, {Duncan}, {Haverkorn}, {Heald}, {Iacobelli}, {Intema}, {Jackson},
  {Jamrozy}, {Jarvis}, {Lakhoo}, {Mevius}, {Miley}, {Morabito}, {Morganti},
  {Nisbet}, {Orr{\'u}}, {Perkins}, {Pizzo}, {Schrijvers}, {Smith}, {Vermeulen},
  {Wise}, {Alegre}, {Bacon}, {van Bemmel}, {Beswick}, {Bonafede}, {Botteon},
  {Bourke}, {Brienza}, {Calistro Rivera}, {Cassano}, {Clarke}, {Conselice},
  {Dettmar}, {Drabent}, {Dumba}, {Emig}, {En{\ss}lin}, {Ferrari}, {Garrett},
  {G{\'e}nova-Santos}, {Goyal}, {G{\"u}rkan}, {Hale}, {Harwood}, {Heesen},
  {Hoeft}, {Horellou}, {Jackson}, {Kokotanekov}, {Kondapally},
  {Kunert-Bajraszewska}, {Mahatma}, {Mahony}, {Mandal}, {McKean}, {Merloni},
  {Mingo}, {Miskolczi}, {Mooney}, {Nikiel-Wroczy{\'n}ski}, {O'Sullivan},
  {Quinn}, {Reich}, {Roskowi{\'n}ski}, {Rowlinson}, {Savini}, {Saxena},
  {Schwarz}, {Shulevski}, {Sridhar}, {Stacey}, {Urquhart}, {van der Wiel},
  {Varenius}, {Webster}, \& {Wilber}}]{2019A&A...622A...1S}
{Shimwell}, T.~W., {Tasse}, C., {Hardcastle}, M.~J., {et~al.} 2019, \aap, 622,
  A1

\bibitem[{{Smirnov}(2011{\natexlab{a}})}]{2011A&A...527A.106S}
{Smirnov}, O.~M. 2011{\natexlab{a}}, \aap, 527, A106

\bibitem[{{Smirnov}(2011{\natexlab{b}})}]{2011A&A...527A.107S}
{Smirnov}, O.~M. 2011{\natexlab{b}}, \aap, 527, A107

\bibitem[{{Smirnov}(2011{\natexlab{c}})}]{2011A&A...527A.108S}
{Smirnov}, O.~M. 2011{\natexlab{c}}, \aap, 527, A108

\bibitem[{{Smirnov}(2011{\natexlab{d}})}]{2011A&A...531A.159S}
{Smirnov}, O.~M. 2011{\natexlab{d}}, \aap, 531, A159

\bibitem[{{Smirnov} \& {Tasse}(2015)}]{2015MNRAS.449.2668S}
{Smirnov}, O.~M. \& {Tasse}, C. 2015, \mnras, 449, 2668

\bibitem[{{Tasse} {et~al.}(2018){Tasse}, {Hugo}, {Mirmont}, {Smirnov},
  {Atemkeng}, {Bester}, {Hardcastle}, {Lakhoo}, {Perkins}, \&
  {Shimwell}}]{2018A&A...611A..87T}
{Tasse}, C., {Hugo}, B., {Mirmont}, M., {et~al.} 2018, \aap, 611, A87

\bibitem[{{van Haarlem} {et~al.}(2013){van Haarlem}, {Wise}, {Gunst}, {Heald},
  {McKean}, {Hessels}, {de Bruyn}, {Nijboer}, {Swinbank}, {Fallows},
  {Brentjens}, {Nelles}, {Beck}, {Falcke}, {Fender}, {H{\"o}randel},
  {Koopmans}, {Mann}, {Miley}, {R{\"o}ttgering}, {Stappers}, {Wijers},
  {Zaroubi}, {van den Akker}, {Alexov}, {Anderson}, {Anderson}, {van Ardenne},
  {Arts}, {Asgekar}, {Avruch}, {Batejat}, {B{\"a}hren}, {Bell}, {Bell}, {van
  Bemmel}, {Bennema}, {Bentum}, {Bernardi}, {Best}, {B{\^\i}rzan}, {Bonafede},
  {Boonstra}, {Braun}, {Bregman}, {Breitling}, {van de Brink}, {Broderick},
  {Broekema}, {Brouw}, {Br{\"u}ggen}, {Butcher}, {van Cappellen}, {Ciardi},
  {Coenen}, {Conway}, {Coolen}, {Corstanje}, {Damstra}, {Davies}, {Deller},
  {Dettmar}, {van Diepen}, {Dijkstra}, {Donker}, {Doorduin}, {Dromer}, {Drost},
  {van Duin}, {Eisl{\"o}ffel}, {van Enst}, {Ferrari}, {Frieswijk}, {Gankema},
  {Garrett}, {de Gasperin}, {Gerbers}, {de Geus}, {Grie{\ss}meier}, {Grit},
  {Gruppen}, {Hamaker}, {Hassall}, {Hoeft}, {Holties}, {Horneffer}, {van der
  Horst}, {van Houwelingen}, {Huijgen}, {Iacobelli}, {Intema}, {Jackson},
  {Jelic}, {de Jong}, {Juette}, {Kant}, {Karastergiou}, {Koers}, {Kollen},
  {Kondratiev}, {Kooistra}, {Koopman}, {Koster}, {Kuniyoshi}, {Kramer},
  {Kuper}, {Lambropoulos}, {Law}, {van Leeuwen}, {Lemaitre}, {Loose}, {Maat},
  {Macario}, {Markoff}, {Masters}, {McFadden}, {McKay-Bukowski}, {Meijering},
  {Meulman}, {Mevius}, {Middelberg}, {Millenaar}, {Miller-Jones}, {Mohan},
  {Mol}, {Morawietz}, {Morganti}, {Mulcahy}, {Mulder}, {Munk}, {Nieuwenhuis},
  {van Nieuwpoort}, {Noordam}, {Norden}, {Noutsos}, {Offringa}, {Olofsson},
  {Omar}, {Orr{\'u}}, {Overeem}, {Paas}, {Pand ey-Pommier}, {Pandey}, {Pizzo},
  {Polatidis}, {Rafferty}, {Rawlings}, {Reich}, {de Reijer}, {Reitsma},
  {Renting}, {Riemers}, {Rol}, {Romein}, {Roosjen}, {Ruiter}, {Scaife}, {van
  der Schaaf}, {Scheers}, {Schellart}, {Schoenmakers}, {Schoonderbeek},
  {Serylak}, {Shulevski}, {Sluman}, {Smirnov}, {Sobey}, {Spreeuw}, {Steinmetz},
  {Sterks}, {Stiepel}, {Stuurwold}, {Tagger}, {Tang}, {Tasse}, {Thomas},
  {Thoudam}, {Toribio}, {van der Tol}, {Usov}, {van Veelen}, {van der Veen},
  {ter Veen}, {Verbiest}, {Vermeulen}, {Vermaas}, {Vocks}, {Vogt}, {de Vos},
  {van der Wal}, {van Weeren}, {Weggemans}, {Weltevrede}, {White}, {Wijnholds},
  {Wilhelmsson}, {Wucknitz}, {Yatawatta}, {Zarka}, {Zensus}, \& {van
  Zwieten}}]{2013A&A...556A...2V}
{van Haarlem}, M.~P., {Wise}, M.~W., {Gunst}, A.~W., {et~al.} 2013, \aap, 556,
  A2

\bibitem[{{Varenius} {et~al.}(2016){Varenius}, {Conway}, {Marti-Vidal},
  {Aalto}, {Barcos-Munoz}, {Koenig}, {Perez-Torres}, {Deller}, {Moldon},
  {Gallagher}, {Yoast-Hull}, {Horellou}, {Morabito}, {Alberdi}, {Jackson},
  {Beswick}, {Carozzi}, {Wucknitz}, \&
  {Ramirez-Olivencia}}]{2016yCat..35930086V}
{Varenius}, E., {Conway}, J.~E., {Marti-Vidal}, I., {et~al.} 2016, VizieR
  Online Data Catalog, J/A+A/593/A86

\bibitem[{{Varenius} {et~al.}(2015){Varenius}, {Conway}, {Mart{\'\i}-Vidal},
  {Beswick}, {Deller}, {Wucknitz}, {Jackson}, {Adebahr}, {P{\'e}rez-Torres},
  {Chy{\.z}y}, {Carozzi}, {Mold{\'o}n}, {Aalto}, {Beck}, {Best}, {Dettmar},
  {van Driel}, {Brunetti}, {Br{\"u}ggen}, {Haverkorn}, {Heald}, {Horellou},
  {Jarvis}, {Morabito}, {Miley}, {R{\"o}ttgering}, {Toribio}, \&
  {White}}]{2015A&A...574A.114V}
{Varenius}, E., {Conway}, J.~E., {Mart{\'\i}-Vidal}, I., {et~al.} 2015, \aap,
  574, A114

\bibitem[{{Zarka} {et~al.}(2012){Zarka}, {Girard}, {Tagger}, \&
  {Denis}}]{2012sf2a.conf..687Z}
{Zarka}, P., {Girard}, J.~N., {Tagger}, M., \& {Denis}, L. 2012, in SF2A-2012:
  Proceedings of the Annual meeting of the French Society of Astronomy and
  Astrophysics, ed. S.~{Boissier}, P.~{de Laverny}, N.~{Nardetto}, R.~{Samadi},
  D.~{Valls-Gabaud}, \& H.~{Wozniak}, 687--694

\bibitem[{{Zarka} {et~al.}(2015){Zarka}, {Tagger}, {Denis}, {Girard},
  {Konovalenko}, {Atemkeng}, {Arnaud}, {Azarian}, {Barsuglia}, {Bonafede},
  {Boone}, {Bosma}, {Boyer}, {Branchesi}, {Briand}, {Cecconi}, {Célestin},
  {Charrier}, {Chassande-Mottin}, {Coffre}, {Cognard}, {Combes}, {Corbel},
  {Courte}, {Dabbech}, {Daiboo}, {Dallier}, {Dumez-Viou}, {El Korso},
  {Falgarone}, {Falkovich✝}, {Ferrari}, {Ferrari}, {Ferrière}, {Fevotte},
  {Fialkov}, {Fullekrug}, {Gérard}, {Grießmeier}, {Guiderdoni}, {Guillemot},
  {Hessels}, {Koopmans}, {Kondratiev}, {Lamy}, {Lanz}, {Larzabal}, {Lehnert},
  {Levrier}, {Loh}, {Macario}, {Maintoux}, {Martin}, {Mary}, {Masson},
  {Miville-Deschenes}, {Oberoi}, {Panchenko}, {Pandey-Pommier}, {Petiteau},
  J.-L., {Revenu}, {Rible}, {Richard}, O., {Salomé}, {Semelin}, {Serylak},
  {Smirnov}, {Stappers}, {Taffoureau}, {Tasse}, {Theureau}, {Tokarsky},
  {Torchinsky}, {Ulyanov}, {van Driel}, {Vasylieva}, {Vaubaillon}, {Vazza},
  {Vergani}, {Was}, {Weber}, \& {Zakharenko}}]{icatt2015}
{Zarka}, P., {Tagger}, M., {Denis}, L., {et~al.} 2015, in International
  Conference on Antenna Theory and Techniques (ICATT), Karkiv, Ukraine, 13--18

\end{thebibliography}
